\documentclass[floatfix,twocolumn,showpacs,preprintnumbers,amsmath,amssymb]{revtex4}

\usepackage{graphicx,epsfig,amsfonts}
\usepackage{dcolumn}
\usepackage{psfrag}
\usepackage{concmath,charter}
\usepackage{subfigure}
\begin{document}

\newcommand{\e}[1]{\emph{#1}} 
\newcommand{\avg}[1]{\langle #1 \rangle}
\newcommand{\va}[0]{{\mathbf a}}
\newcommand{\vb}[0]{{\mathbf b}}
\newcommand{\vc}[0]{{\mathbf c}}

\title{Global statistical analysis of the protein homology network}

\author{C.~Miccio}
\email{miccio@mib.infn.it}
\affiliation{
  Dipartimento di Fisica G.Occhialini, Universit\`a di
  Milano--Bicocca and INFN, Sezione di Milano, Piazza della Scienza 3
  - I-20126 Milano, Italy}
\author{T.~Rattei} 
\email{t.rattei@wzw.tum.de} 
\affiliation{
  Department of Genome Oriented Bioinformatics, Technical University
  of Munich, 
  Wissenschaftszentrum 5 Weihenstephan, 85350 Freising,
  Germany }

\date{\today}

\begin{abstract}
  The similarity between protein sequences is a directly and easly
  computed quantity from which to deduce information about their
  evolutionary distance and to detect homologous proteins. The {\emph
    SIMAP} database -- {\emph Similarity Matrix of Proteins} --
  provides a pre-computed similarity matrix covering the similarity
  space formed by about all publicly available amino acid sequences
  from public databases and completely sequenced genomes.  From SIMAP
  we construct the protein homology network, where the proteins are
  the nodes and the links represent homology relationships.  With more
  than $5$ million nodes and about $70 \times 10^9$ edges it is the
  greatest protein homology network ever been builded.  We
  describe the basic features and we perform a global statistical
  analysis of the network. Starting from the Smith-Waterman similarity
  score, we define for each edge a weight $w$ to measure the
  similarity distance between two nodes. Keeping only edges with a
  weigth greater than a minimal $\bar w$, and by varying $\bar w$ we
  build a family of networks with different degree of similarity. We
  investigate the distribution of connected components (clusters) of
  the networks at different $\bar w$ and in particular we find a
  behaviour similar to a phase transition guided by the formation of a
  giant component. Moreover we study selected sequence features and
  protein domains of protein pairs that connect different clusters in
  the networks at different level of similarity.  We observed
  specific, non-random distributions of the protein features and
  domains for proteins connecting clusters at certain weight
  intervals.
\end{abstract}

\maketitle

\section{Background}

The number of known proteins is rapidly growing and the sequence of
amino acids is, at the moment, the main source of information for many
new proteins which still have unidentified functions. Protein sequence
analysis, and more specifically, the analysis of similarities among
protein sequences, is therefore the basis of studies trying to
understand protein evolutionary processes or to detect unknown
biological functions of new proteins. Proteins with similar sequences
can be found in different organisms and in a single organism
\footnote{Due to duplication and shuffling of coding segments in the
  akno DNA during the evolution.}, \cite{revEvol}. By means of the
degree of similarity obtained by a pairwise sequence comparison it is
possible to deduce information about their evolutionary distance.
Specifically, two proteins are homologous if they evolved from a
common ancestral protein sequence and, in most cases, they have also
the same, or very similar, biological function. Homology can be
deduced from statistically significant sequence similarities. However,
new sequences often have only weak similarities to known proteins, and
single similarities search are insufficient to assign validated
properties of characterized proteins to new sequences. Instead a graph
formed by all-against-all comparisons of a large amount of
protein-data could become useful. This is the case of {\bf SIMAP} --
\e{Similarity Matrix of Proteins} -- a database containing the
similarity space formed by almost all amino acid sequences, with
nearly 5.5 million non-redundant protein sequences drawn from
completely sequenced genomes and public database. Moreover,
pre-calculated similarity space allows very rapid access to
significant hits of interest and prevents time-consuming
re-computation. The algorithm that precomputes the sequences
similarities is based on the FASTA heuristic. First it compares
low-complexity masked proteins using FASTA and then it recalculates
the hits found using non-masked sequences and the Smith-Waterman
algorithm. In both phases of the alignment process the BLOSUM50 amino
acids substitution matrix is used. For each hit the Smith-Waterman
score, the identity, the gapped identity, the overlap and the start
and the stop coordinates of the alignment in
both proteins are stored. For more details see \cite{simap}.\\
Graphs formed by all-against-all sequence comparisons can be used to
derive inheritance patterns of proteins, to reconstruct the
evolutionary relationships between proteins and to classify them into
protein families by looking for dense clusters disconnected from the
rest of the network. To date, this approach has been carefully
evaluated by case studies targeted at selected protein families
\cite{phn}, but a global analysis of the complete homology network
formed by all publicly available proteins has not been published. The
aim of this work is to analyze global and local properties of the
graph forming the homology network.

\section{SIMAP graph representation}

The information contained in the Simap database can be reorganized by
means of a weighted graph representation, $G(V, E, w)$, where $V$ is
the set of nodes, $E$ the set of edges, and $w$ a weight function on
the edges: $w : E \to [0,1]$. Each node, $\va \in V$, represents a
protein sequence and each edge, $e = \{ \va,\vb \} \in E$ between two
nodes $\va$, $\vb$ represents the stored alignment between the
respective protein sequences\footnote{For simplicity we will use the
  same notation to point graphs's nodes and database's proteins.}.  In
this way an undirected weighted graph can be obtained, since the
symmetry of the alignment procedure leads to undirected edges and the
score of the alignment allows the assignment of a suitable weight to
every edge. (Despite the possibility of making an alignment between a
protein sequence and itself, self-edges are not considered). More
specifically if $s(\va,\vb)$ is the Smith-Waterman (SW) optimal score
obtained with the FASTA algorithm between sequence $\va$ and $\vb$, a
suitable weight $w(\va,\vb) \in [0,1]$ for the edge $e = \{ \va,\vb
\}$ can be defined as follow:
\begin{equation}
\label{eq:weight} 
  w(\va,\vb) = \frac{s(\va,\vb)}{ \sqrt{ \; s(\va,\va) \;
      s(\vb,\vb)}}, 
\end{equation}
From $w(\va,\vb)$ one could define a distance function as $d(\va,\vb)
= 1 - w(\va,\vb)\;$, whose values are in $[0, 1]$ as distance function
usually defined on linear spaces. $d$ should satisfy positivity, null
and simmetry properties for all pairs of sequence proteins and also
the triangular inequality which is fully satisfied for the BLOSUM50
matrix.

\section{Polishing procedure}
Strictly speaking, the set of all protein sequences of the Simap
database is not a good space over which to define the distance measure
$d$. There are, in fact, $1538$ pairs of sequences that have distance
equal to zero, although they are classified with a different sequence
id. However, they differ only in the presence of one or two $'$X$'$ in
their amino acid sequence annotation, where $'$X$'$ is the standard
symbol for an unknown amino acid residue in a protein sequence.  It is
therefore natural to decide to knock out, for each of these pairs of
sequences, the one that has the $'$X$'$ in the sequence; this
procedure entails the removal, in the graph representation, of all
edges connected to the removed nodes. Another improvment for database
consistency is the checking of symmetry of all edges: every time, a
direct edge is found, the inverse relation, if absent, is added.

As a final result of these manipulations, a graph with $V = 5,489,907$
nodes and $E = 69,500,722,050$ edges can be constructed.

Over the polished Simap protein sequences space the distance $d = 1 -
w(\va,\vb)\;$ fails the triangular inequality over few cases (around
$\approx 0.2 \%$ of triangles). However redefining, for istance,

\begin{equation}
\label{eq:distance} 
d(\va,\vb) =\sqrt{1 - w(\va,\vb)}, 
\end{equation}
we have that the triangle inequality is satisfied for all triples of
linked proteins and (\ref{eq:distance}) has all properties required
for a \e{distance measure}.

\section{Characterization of Simap protein space}

In the Simap database, protein sequences come from $104,560$ different
species.  There are, in particular, $3$ species (\e{Homo sapiens},
\e{Arabidopsis thaliana}, \e{Rice plants}) with more than $100,000$
protein sequences and $72$ with more than $10,000$.

\begin{table}[!htb]
  \begin{center}
      \begin{tabular}{|c|c|c|}
        \vspace{-10pt} & & \\ \hline \it{kingdoms} & & \it{number of species} \\
        \vspace{-10pt} & & \\ \hline
        bacteria &                &  $11,130$    \\ \hline
        viruses  &  viruses       &  $13,708$    \\
        &  phages        &  $923$     \\ \hline
        plants   &                &  $31,232$   \\ \hline
        animalia & invertebrates  &  $25,951$   \\ 
        & vertebrates    &  $19,341$   \\
        & (rodents)      &  $(1,474)$  \\
        & (mammals)      &  $(1,854)$  \\
        & (primates)     &  $(393)$   \\ \hline	         
        environmental samples  &   &  $1,453$    \\ \hline
        synthetic              &   &  $822$      \\ \hline
      \end{tabular} 
      \caption{\label{tab1} \small Number of species for each
        kingdom.}
  \end{center}
\end{table}

\begin{figure}[!htb]
  \begin{center}
    \includegraphics[height=0.36\textwidth,angle=270]{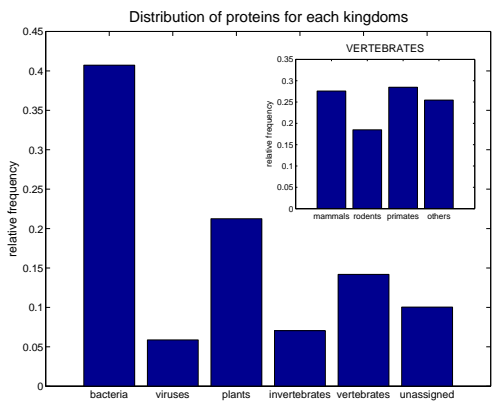}
    \caption{\label{fig1} {\small Distribution of
        proteins for each kingdom. The little graph shows the
        distribution within vertebrates.}}
  \end{center}
\end{figure}

A coarse subdivision of all species is shown in
Table~\ref{tab1}; it separates species in five (non-standard)
main kingdoms: bacteria, viruses, plants, invertebrates (animalia) and
vertebrates (animalia). The classification reveals the presence of
very many different animalia species, but only eight of these species
are present with their complete genome (the other animalia proteins
were imported from multiple species databases).
Figure~\ref{fig1} shows the protein distribution for
each kingdom. There is also a high number ($546,439$) of unassigned
protein sequences.\footnote{These sequences come from databases:
  \e{PDB proteins}, \e{mips non-redundant protein database},
  \e{UNIPROT SWISSPROT}, \e{UNIPROT-TrEMBL}, \e{PFAM sequences},
  \e{Eukaryotic signature proteins.}}.

\subsection{Length and self-similarity distribution}

\begin{figure}[!htb]
  \includegraphics[height=0.70\textwidth]{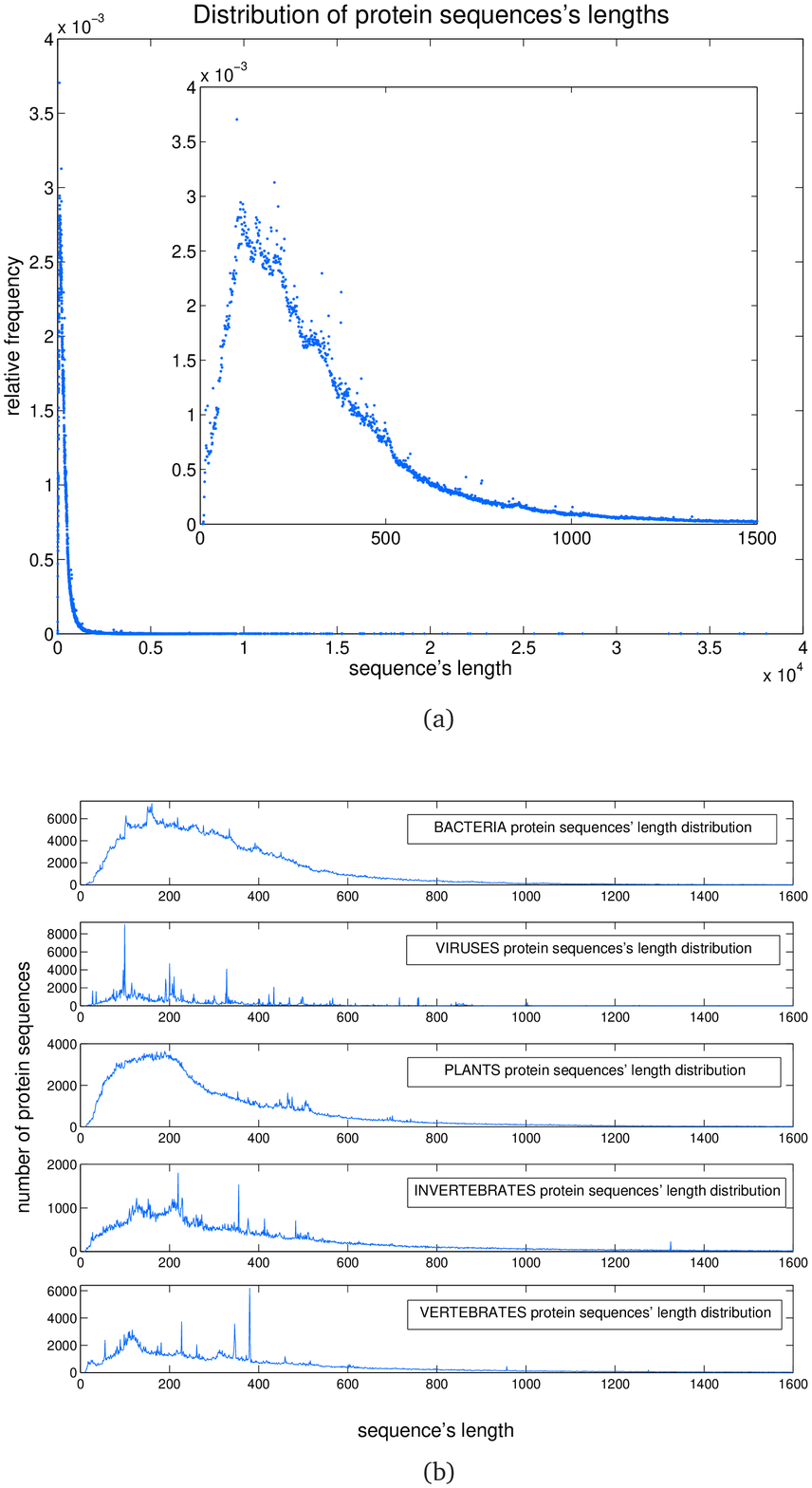}
  \caption{\label{fig2}{\small (a) Distribution of protein sequences'
      lengths. In the inner boxe an enlargement of the distribution is
      shown. (b) Length distributions of protein sequences which
      belong to \e{bacteria} ($\avg{l} = 316.9$, $l_{max} = 36805$),
      \e{viruses} ($\avg{l} = 273.9$,$l_{max} = 7312$ ), \e{plants}
      ($\avg{l} = 314.5$, $l_{max} = 20925$), \e{invertebrated}
      ($\avg{l} = 416.1$, $l_{max} = 23015$), \e{vertebrated}
      ($\avg{l} = 397.1$, $l_{max} = 38031$).}}
\end{figure}

The protein sequences space is characterized by the length
distribution shown in Figure~\ref{fig2}a and in Figure~\ref{fig2}b we
give the length distributions for sequences belonging to bacteria,
viruses, plants, vertebrates and invertebrates.

\begin{figure}[!htb]
  \vspace{0.2cm}
  \includegraphics[height=0.36\textwidth,angle=270]{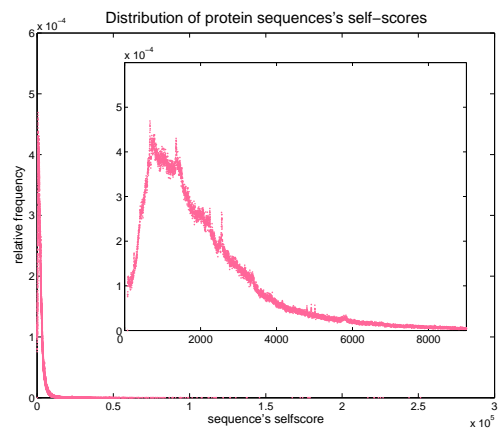}
  \caption{\label{fig3}{\small Distribution of protein sequences'
      self-scores. In the inner boxe an enlargement of the
      distribution is shown. }}
\end{figure}

\begin{figure}[!htb]
  \vspace{0.2cm}
  \includegraphics[height=0.36\textwidth,angle=270]{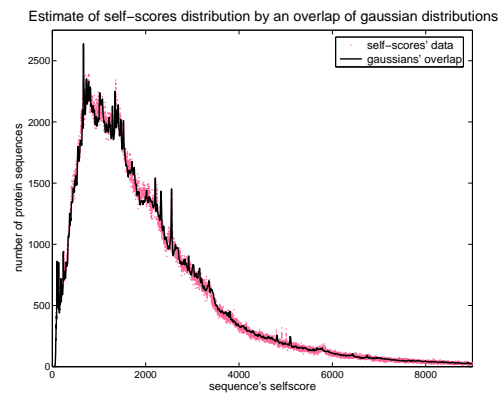}
  \caption{\label{fig4}{\small Distribution of protein sequences'
      self-scores and the curve obtained by an overlap of normal
      distributions opportunely wighted by the protein sequences's
      length distribution are compared.}}
\end{figure}

The \e{self-similarity} \e{score} 's distribution of protein sequence
appears in Figure~\ref{fig3}.  The self-similarity scores distribution
is well reproduced by a mixture of normal distributions, one for each
length entry. The self-similarity score $s(\va,\va)$ of a protein
sequence of length $l$, can be thougth as a sum of $l$ i.i.d.  random
variables, i.e. a sum of the self-similarities scores of random amino
acids. Knowing the amino acids background probabilities\footnote{ The
  values for background distribution of amino acids come from data
  used for the PAM matrix: $\;p_A=0.096;\; p_R=0.034;\; p_N=0.042;\;
  p_D=0.053;\; p_C=0.025;\; p_Q=0.032;\; p_E=0.053;\; p_G=0.090;\;
  p_H=0.034;\; p_I=0.035;\; p_L= 0.084;\; p_K=0.085;\; p_M=0.012;\;
  p_F=0.045;\; p_P=0.041;\; p_S=0.057;\; p_T=0.062;\; p_W=0.012;\;
  p_Y=0.030;\; p_V=0.078$.\\ They can be obtained from \e{{\small
      http://apps.bioneq.qc.ca/twiki/pub/Knowledgebase/PAM/}}
  \e{{\small PAM2.JPG}}} $p_{a}$ and the diagonal values of the
BLOSUM50 score matrix, $B_{aa}$, the self-similarity score of a random
amino acid will follow a normal distribution with mean $\avg{s} =
\sum_a p_a \, B_{aa} \;\; ( \approx 6.727)$ and variance $ \sigma =
\sqrt{\sum_a p_a B_{aa}^2 - \avg{s} ^2} \;\;(\approx 2.067)$.
Self-similarity scores of random amino acid sequences of length $l$
will have a normal distribution $g(l,s)$ with mean $l\,\avg{s}$ and
variance $\sqrt{l\,\sigma^2}$.  Finally, the self-similarity scores
distribution is well approximated by the sum $\sum_{l} g(l,s) f(l)$,
where $f(l)$ is the observed length distribution, Figure~\ref{fig4}.

\subsection{Pairwise similarity distribution}

The SW optimum similarity scores distribution obtained from all FASTA
sequence alignments present a homogeneous cutoff equal to $80$, used
for storing hits in Simap database. It was chosen independently of the
query and database length, but as an optimal compromise between
sensitivity and possibility to store an accessible number of hits,
because of the high number of protein sequences.

\begin{figure}[!htb]
  \includegraphics[height=0.70\textwidth]{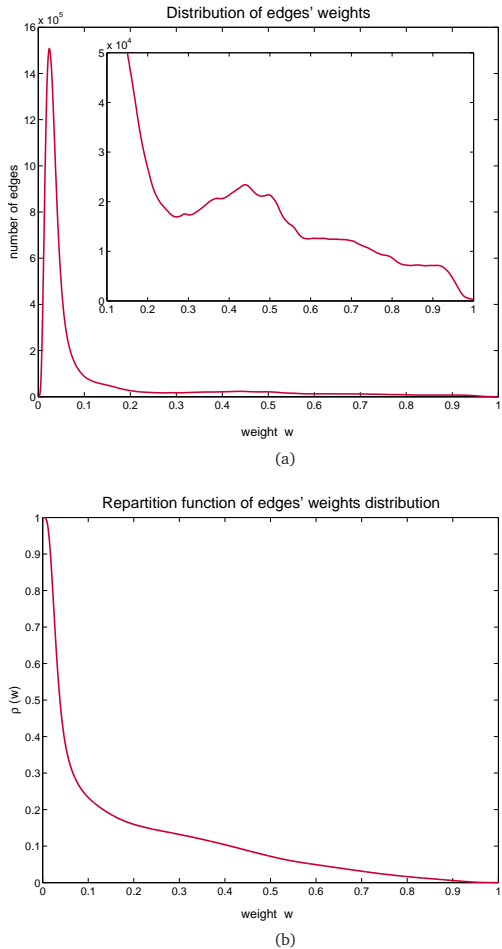}
  \caption{\label{fig5}{\small (a) Distribution of edges' weights $w$.
      In the inner box is shown an enlargement of the distribution
      tail. (b) Repartition function edges' weights distribution.}}
\end{figure}

In Figure~\ref{fig5}a the distribution of weights $w$ is shown, and in
Figure~\ref{fig5}b the corresponding repartition distribution $\rho(w)$. The
values of $\rho(w) \in [0,1]$ represent the fractions of edges which
have weight greater or equal to $w$. From them we see that the major
part of the edges (about $80\%$ of the total number of edges) has a
very low value of $w$ ($\leq 0.2$).
 
\subsection{Coordination and cluster distribution}

Weights $w$ can be used as a parameter to define a collection of
graphs. For a fixed value of $w = \bar{w}$ (or a value of $d = \bar{d}
= \sqrt{1 -\bar{w} }$ ), a graph is built keeping only edges with $w >
\bar{w}$ ($d \le \bar{d}$). For high values of $\bar{w}$, i.e. at
small distances, nodes are linked if, and only if, the corresponding
protein sequences have a high degree of similarity; then it is
reasonable to expect graphs with many small connected components. By
decreasing $\bar{w}`$ values, in other words by also linking proteins
having a lower degree of similarity, graphs with larger connected
components are expected. The graph obtained by considering all
possible edges (by fixing $\bar{w} = 0$) is not the complete graph,
due to the cutoff on the score alignment (there are about $0.1 \%$ of
edges of the corresponding complete graph).

We have built graphs for values of $w$ equal to $0.975$, $0.95$,
$0.925$, $ 0.9$, $0.875$, $0.85$, $0.825$, $0.8$, $0.775$, $0.75$,
$0.725$, $0.7$, $0.675$, $0.65$, $0.625$, $0.6$, $0.575$, $0.55$,
$0.525$, $0.5$, $0.475$, $0.45$, $0.425$, $0.4$ $0.375$, $0.35$,
$0.325$, $0.3$, $0.275$, $0.25$, $0.225$, $0.2$, $0.175$, $0.15$,
$0.125$; $\;$ for each of these values the set of the protein
sequences splits into clusters, i.e. isolated connected components.
Linking proteins that have a greater and greater distance from each
other (decresing $\bar{w}$), clusters merge to form larger clusters,
the number of isolated proteins and the number of components with a
very small size decreases, while the number of clusters of medium and
large size increases.

\begin{figure}[!htb]

    \subfigure[]{\label{fig6a} 
      \includegraphics[width=0.34\textwidth, angle=270]{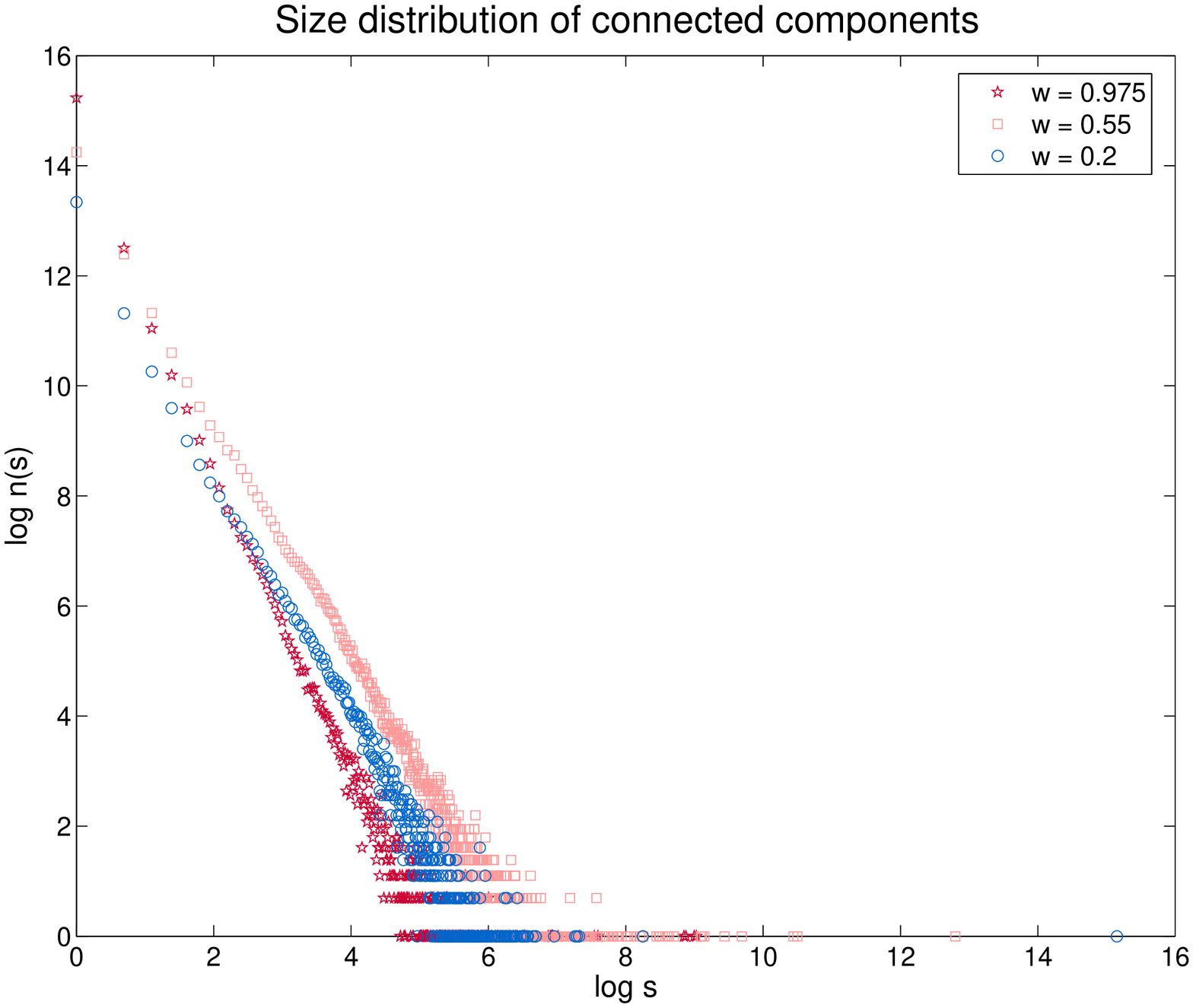}
    } \vspace{-0.4cm}
    \subfigure[]{\label{fig6b}
      \includegraphics[width=0.40\textwidth, height=0.42\textwidth, angle=270]{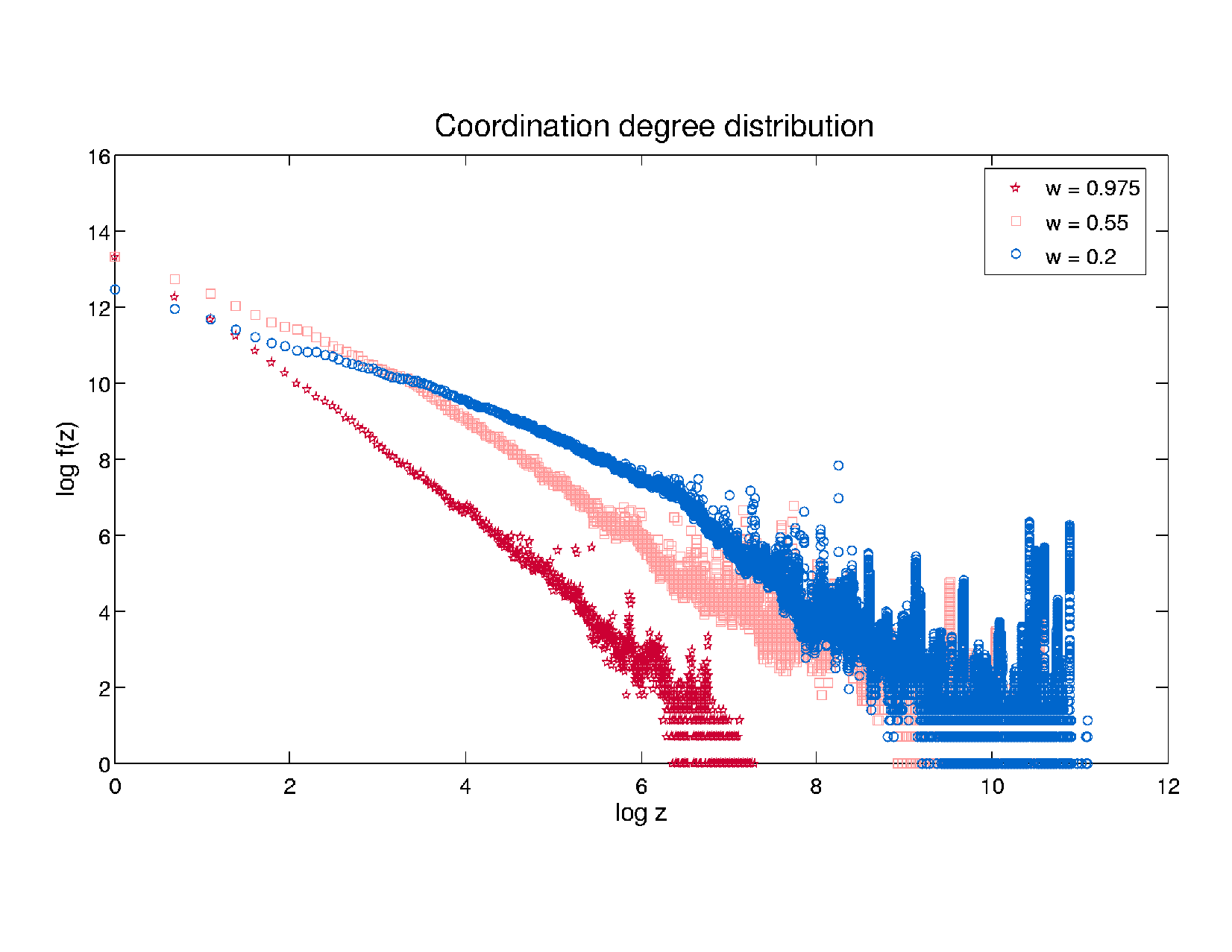}
    }

  \caption{{\small (a) Distribution of size of connected
      components of the protein sequences graph built at $\bar{w} =
      0.975$ (red curve), $\bar{w} = 0.75$ (pink curve) and $\bar{w} =
      0.4$ (blue curve). It is evident that as the $\bar{w}$ value
      decrease the number of connected components with small size
      decreases and the starting region of the power law behaviour
      shifts to higher values of size. (b) Distribution of
      coordination degree of the protein sequences graph built at
      $\bar{w} = 0.975$ (red curve), $\bar{w} = 0.75$ (pink curve) and
      $\bar{w} = 0.4$ (blue curve). As the $\bar{w}$ value decrease
      the number of nodes with coordination degree decreases and the
      starting region of the power law behaviour shifts to higher
      values of coordination degree.}}
\end{figure}

Measuring the (not normalized) cluster distribution, we find that, for
each fixed values of $\bar{w}$, the number of clusters
$n_{\bar{w}}(s)$ of size $s$ follows, in a specific size range, a
power law behaviour, $n_{\bar{w}}(s) \sim s^{-\sigma(\bar{w})}$.
Fitted values of $\sigma(\bar{w})$ and fitting size ranges are
reported in Table~\ref{tab2} and a log-log plot of size
distribution $n_{\bar{w}}(s)$, for three different values of $\bar w$
is shown in Figure~\ref{fig6a}.  Also the (not normalized) coordination
degree distribution $f_{\bar w}(z)$ follows a power law distribution,
$f_{\bar w}(z) \sim z^{-\alpha(\bar w)}$, for each values of
$\bar{w}$. A log-log plot of coordination degree distribution
$f_{\bar{w}}(z)$, for three different values of $\bar w$ is shown in
Figure~\ref{fig6b}.  Fitted values of $\alpha(\bar{w})$ and fitting
coordination degree's ranges are reported in Table~\ref{tab3}.

\begin{table}[!htb]
  \begin{center}
      \begin{tabular}{|c|ccc|} \hline
        \vspace{-10pt} & & & \\ $\bar{w}$ & $\sigma$ & \quad component
        & \quad correlation \\ & & \quad size range & \quad coefficient \\
        \vspace{-10pt} & & & \\ \hline
        \vspace{-10pt} & & & \\
        $\;$ $0.95$ $\;$   &   $\;$ $2.70$  &  $10 - 60$  &  $-0.995$ \\
        $\;$ $0.90$ $\;$    &   $\;$ $2.70$  &  $10 - 60$  &  $-0.996$ \\
        $\;$ $0.85$ $\;$    &   $\;$ $2.69$  &  $10 - 60$  &  $-0.994$ \\
        $\;$ $0.80$ $\;$    &   $\;$ $2.62$  &  $10 - 80$  &  $-0.996$ \\
        $\;$ $0.75$ $\;$    &   $\;$ $2.52$  &  $10 - 80$  &  $-0.996$ \\
        $\;$ $0.70$ $\;$    &   $\;$ $2.40$  &  $10 - 80$  &  $-0.996$ \\
        $\;$ $0.65$ $\;$    &   $\;$ $2.32$  &  $10 - 100$  &  $-0.997$ \\
        $\;$ $0.60$ $\;$    &   $\;$ $2.21$  &  $10 - 100$  &  $-0.996$ \\
        $\;$ $0.55$ $\;$    &   $\;$ $2.17$  &  $10 - 100$  &  $-0.996$ \\
        $\;$ $0.50$ $\;$    &   $\;$ $2.07$  &  $10 - 100$  &  $-0.997$ \\
        $\;$ $0.45$ $\;$    &   $\;$ $2.01$  &  $10 - 100$  &  $-0.997$ \\
        $\;$ $0.40$ $\;$    &   $\;$ $2.00$  &  $10 - 100$  &  $-0.996$ \\
        $\;$ $0.35$ $\;$    &   $\;$ $1.98$  &  $10 - 100$  &  $-0.997$ \\
        $\;$ $0.30$ $\;$    &   $\;$ $1.98$  &  $10 - 100$  &  $-0.997$ \\
        $\;$ $0.25$ $\;$    &   $\;$ $2.01$  &  $10 - 100$  &  $-0.996$ \\ \hline
      \end{tabular}
    \caption{\label{tab2} \small Fitting values of exponent
      $\sigma$ of the power law distribution of connected components
      for selected values of $\bar{w}$. For each fitting the size
      range and its correlation coefficient are reported.}
\end{center}
\end{table}

\begin{table}[!htb]
  \begin{center}
    \begin{tabular}{|c|c|c|ccc|} \hline
      \vspace{-10pt} & & & &\\ 
      
      $\bar{w}$ &  $\avg{z}$ & max $z$ & $\alpha$ & \quad coordination & \quad
      correlation \\
      
      & & & & \quad degree range & \quad coefficient \\
      \vspace{-10pt} & & & & &\\ \hline
      \vspace{-10pt} & & & & &\\
      
      $0.95$   & $14.4$    & $5735$  & $1.59$  & $25 - 100$    &  $-0.990$ \\
      &           &         & $1.46$  &  $100 - 500$  &  $-0.953$ \\ \hline
      
      $0.90$   & $73.1$    & $10794$  & $1.58$  & $25 - 100$   &  $-0.988$ \\
      &           &          & $1.51$  & $100 - 500$  &  $-0.939$ \\ \hline
        
        $0.85$   & $138.3$   & $16500$  & $1.68$  & $25 - 100$   &  $-0.993$ \\
        &           &          & $1.42$  & $100 - 800$  &  $-0.964$ \\ \hline
        
        $0.80$   & $207.2$   & $ 23726$ & $1.73$  & $25 - 100$   &  $-0.994$ \\
        &           &          & $1.29$  & $100 - 800$  &  $-0.941$ \\ \hline
        
        $0.75$   & $294.0$   & $33265$  & $1.79$  & $25 - 100$   &  $-0.997$ \\
        &           &          & $1.22$  & $100 - 1000$ &  $-0.956$ \\ \hline
        
        $0.70$   & $395.3$   & $35202$  & $1.74$  & $25 - 100$   &  $-0.996$ \\
        &           &          & $1.28$  & $100 - 1000$ &  $-0.946$ \\ \hline	
        
        $0.65$   & $507.8$   & $36333$  & $1.71$  & $25 - 100$   &  $-0.998$ \\
        &           &          & $1.39$  & $100 - 1000$ &  $-0.950$ \\ \hline
        
        $0.60$   & $622.3$   & $37729$  & $1.63$  & $25 - 100$   &  $-0.999$ \\
        &           &          & $1.32$  & $100 - 1500$ &  $-0.930$ \\ \hline
        
        $0.55$   & $745.3$   & $41871$  & $1.54$  & $25 - 100$   &  $-0.998$ \\
        &           &          & $1.44$  & $100 - 1500$ &  $-0.927$ \\ \hline
        
        $0.50$   & $911.7$   & $49895$  & $1.44$  & $25 - 100$   &  $-0.998$ \\
        &           &          & $1.56$  & $100 - 2000$ &  $-0.944$ \\ \hline
        
        $0.45$   & $1108.1$  & $51309$  &  $1.38$  & $25 - 100$  &  $-0.998$ \\
        &           &          &  $1.62$  & $100 - 2000$&  $-0.951$ \\ \hline
        
        $0.40$   & $1314.2$  & $51956$  & $1.28$  & $25 - 100$   &  $-0.998$ \\
        &           &          & $1.67$  & $100 - 2500$ &  $-0.946$ \\ \hline
        
        $0.35$   & $1501.9$  & $52513$  & $1.19$  & $25 - 100$   &  $-0.998$ \\
        &           &          & $1.72$  & $100 - 2500$ &  $-0.961$ \\ \hline
        
        $0.30$   & $1668.9$  & $60722$  & $1.08$  & $25 - 100$   &  $-0.997$ \\
        &           &          & $1.74$  & $100 - 3000$ &  $-0.969$ \\ \hline	
        
        $0.25$   & $1826.2$  & $64781$  & $0.97$  & $25 - 100$   &  $-0.997$ \\
        &           &          & $1.78$  & $100 - 3000$ &  $-0.969$ \\ \hline	
      \end{tabular} 
    \caption{\label{tab3} \small Fitting values of exponent $\alpha$ of
      the power law distribution of coordination degree for selected
      values of $\bar{w}$. We compute two linear fittings different in the
      choice of fitting range of coordination degree. For each fitting the
      range of coordination degree and its correlation coefficient are
      reported. In the second column the average degree is shown; the
      third column gives the maximum value of the coordination degree. }
  \end{center}
\end{table}

\section{Comparison with generalized random graphs}

It would be interesting to compare these behaviours with that of a
model of random graphs. It is well known that, in the classical model,
random graphs (where every pair of nodes is chosen to be an edge with
probability $p$, as introducede by Erd\"os-R\'enyi
\cite{erdos_renyi}), have the same expected coordination degree at
every node, so they are characterized by a poissonian coordination
degree distribution with mean value $\avg{z} \sim p V$. Futhermore, as
soon as $\avg{z}$ assume a value greater than $1$, a giant connected
component appears, that is a component whose size is much greater than
the size of all other components, and that represents an important
fraction of all graph's nodes.

A better theorical comparison model could be represented by
generalized random graphs endowed with a specific degree-distribution.
These can be generated via the Monte-Carlo algorithm (following the
work in \cite{burda} of Burda et al.). In particular, starting from a
random graph of $V$ nodes and $E$ edges, making local graph
transformations which leave the number of nodes and the number of
edges constant and accepting them with a probability which depends on
the desired equilibrium degree distribution (Metropolis algorithm), we
have generated a collection of random graphs with the same
coordination degree distribution and the same average degree as some
of our protein sequences graphs.

\begin{figure}[!htb]
  \includegraphics[height=0.36\textwidth,angle=270]{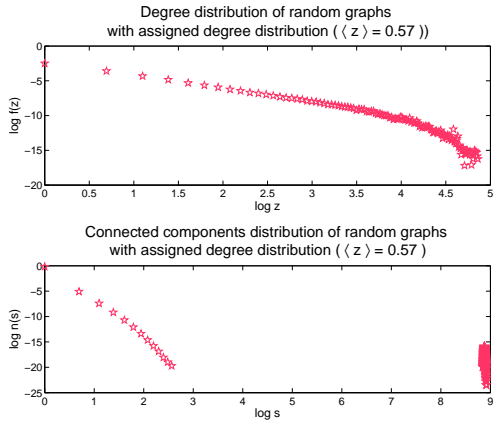}
  \caption{\label{fig7}{\small Top: coordination degree distribution
      of the collection of random graphs generated via Monte-Carlo
      algorithm fixing the equilibrium degree distribution equal to
      that one observed in the protein sequences graph at $\bar{w} =
      0.99$ and fixing the average degree equal to $\avg{z} = 0.57$.
      Bottom: size distribution of connected components of the random
      graphs.}}
\end{figure}

For each of them we observe a fundamentally different
distribution of connected components in the protein sequences graphs
and in the random graphs. In the latter model the power law behaviour
is absent, while there is a always a dominant giant connected
component, much larger than the many other small components, whose
size distribution decreases exponentially (See Figure~\ref{fig7}).

\begin{figure}[!htb]
  \includegraphics[height=0.36\textwidth,angle=270]{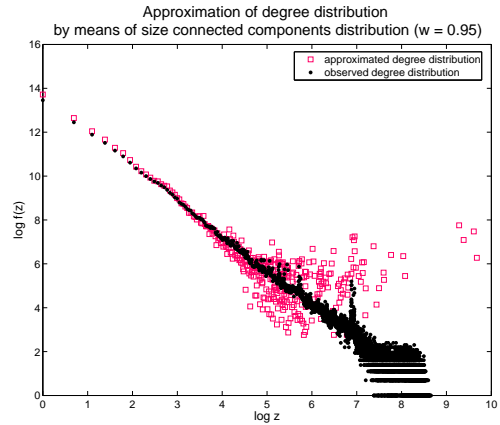}
  \caption{\label{fig8}{\small Observed degree distribution (black
      curve) and the approximated degree distribution (red curve)
      obtained by means of $n(s)$ of the graph at $\bar{w} = 0.95$.}}
\end{figure}

By comparison, in the Simap protein sequences space the coordination
degree distribution $f_{\bar w}(z)$ and the connected component
distribution $n_{\bar w}(s)$ are strongly correlated. The former, for
example, can be reproduced quite well by means of $n_{\bar w}(s)$. Let
the index $i$ label all connected components and let us consider all
possible edges between nodes belonging to a connected components of
size $s_i$; then the cluster would be a complete subgraph and all its
$s_i$ nodes would have coordination degree equal to $z_i = s_i-1$. If
this were true for all connected components then all clusters would be
complete subgraphs and we would expect a coordination degree
distribution equal to $f_{\bar w}(z) \sim ( s \; n_{\bar w}(s) ) |_{s
  = z + 1} $. In our graphs, although complete connected components
are present, the majority of clusters have only a high average degree
distribution, not equal to its size minus one, as in complete graphs.
However let's consider a component with size $s_i$ and a number of
edges equal to $m_i$; the quantity $\Delta_i = \frac{2 m_i}{s_i
  (s_i-1)}$ represents the fraction of edges that are present in the
$i$-th component respect to the number of edges that would be present
if the component were a complete subgraph (i.e. $s_i (s_i-1)/2$).
Introducing $\Delta_i$ as a measure of edges' density for each
component we can approximate the coordination degree distribution
$f_{\bar w}(z)$ by means of the size connected component distribution
$n_{\bar w}(s)$ too. Specifically we find that the coordination degree
distribution behaves like $f_{\bar w}(z) \sim \bar{\Delta}(z+1) \;
(z+1) \; n_{\bar w}(z+1) $, where $\bar{\Delta}(s)$ is the edges'
density averaged over all components of size $s$: $\bar{\Delta}(s) =
\frac{\sum_{i} \delta_{s_i, s} \Delta_i}{\sum_{i} \delta_{s_i, s}}$.
Figure~\ref{fig8} shows both the observed degree distribtution and the
approximated degree distribution obtained by means of $n(s)$ of the
graph at $\bar{w} = 0.95$.

\section{Giant component}

\begin{figure}[!htb]
  \includegraphics[height=0.70\textwidth]{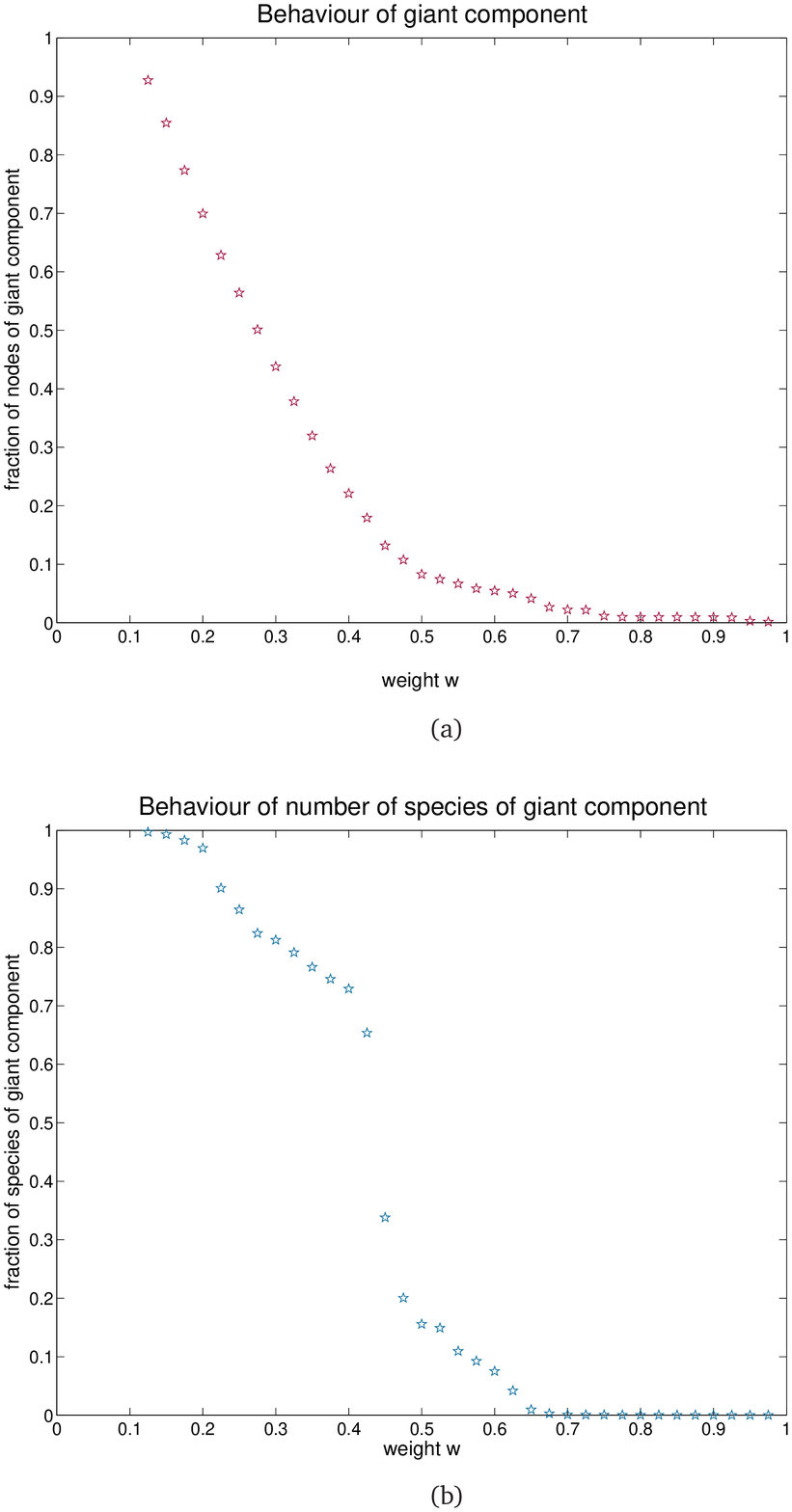}
  \caption{\label{fig9}{\small (a) Fraction of nodes belonging to the
      largest cluster for each value of $\bar{w}$. (b) Fraction of
      species present in the largest cluster for each value of
      $\bar{w}$.}}
\end{figure}

An interesting phenomenon occurs when $\bar{w}$ value decrease; we see
the formation of the giant component.  In Figure~\ref{fig9}a the
behaviour of the fraction of nodes belonging to the largest component
is shown.

\begin{table*}[!htb]
  \begin{center}
  \begin{tabular}{|c|c|c|c|c|c|c|c|c|}\hline 
    \vspace{-7pt} & & & & & & & & \\
    
    $\bar{w}$ & $\bar{d}$ & size & bacteria & viruses & plants &
    invertebrates & vertebrates & number of different species \\
    
    \vspace{-7pt} & & & & & & & & \\ \hline
    
    $0.975$ & $0.1581$ & $8322$ & $0.000$ & $1.000$ & $0.000$ & $0.000$ & $0.000$ & $4$ \\ 
    $0.950$ & $0.2236$ & $15955$ & $0.000$ & $1.000$ & $0.000$ & $0.000$ & $0.000$ & $4$ \\ 
    $0.925$ & $0.2739$ & $47687$ & $0.000$ & $1.000$ & $0.000$ & $0.000$ & $0.000$ & $10$ \\ 
    $0.900$ & $0.3162$ & $50729$ & $0.000$ & $1.000$ & $0.000$ & $0.000$ & $0.000$ & $14$ \\ 
    $0.875$ & $0.3536$ & $51028$ & $0.000$ & $1.000$ & $0.000$ & $0.000$ & $0.000$ & $14$ \\ 
    $0.850$ & $0.3873$ & $51405$ & $0.000$ & $1.000$ & $0.000$ & $0.000$ & $0.000$ & $14$ \\ 
    $0.825$ & $0.4183$ & $51969$ & $0.000$ & $1.000$ & $0.000$ & $0.000$ & $0.000$ & $29$ \\ 
    $0.800$ & $0.4472$ & $52097$ & $0.000$ & $1.000$ & $0.000$ & $0.000$ & $0.000$ & $29$ \\ 
    $0.775$ & $0.4743$ & $52881$ & $0.000$ & $1.000$ & $0.000$ & $0.000$ & $0.000$ & $29$ \\ 
    $0.750$ & $0.5000$ & $63003$ & $0.000$ & $1.000$ & $0.000$ & $0.000$ & $0.000$ & $60$ \\ 
    $0.725$ & $0.5244$ & $118777$ & $0.000$ & $1.000$ & $0.000$ & $0.000$ & $0.000$ & $67$ \\ 
    $0.700$ & $0.5477$ & $120974$ & $0.000$ & $0.999$ & $0.000$ & $0.000$ & $0.000$ & $106$ \\ 
    $0.675$ & $0.5701$ & $145278$ & $0.002$ & $0.997$ & $0.000$ & $0.000$ & $0.000$ & $302$ \\ 
    $0.650$ & $0.5916$ & $224310$ & $0.002$ & $0.749$ & $0.001$ & $0.000$ & $0.248$ & $988$ \\ 
    $0.625$ & $0.6124$ & $272426$ & $0.014$ & $0.662$ & $0.010$ & $0.007$ & $0.306$ & $4384$ \\ 
    $0.600$ & $0.6325$ & $297280$ & $0.028$ & $0.643$ & $0.015$ & $0.011$ & $0.303$ & $7854$ \\ 
    $0.575$ & $0.6519$ & $318472$ & $0.032$ & $0.613$ & $0.027$ & $0.015$ & $0.313$ & $9668$ \\ 
    $0.550$ & $0.6708$ & $362379$ & $0.047$ & $0.554$ & $0.035$ & $0.024$ & $0.341$ & $11437$ \\ 
    $0.525$ & $0.6892$ & $404788$ & $0.049$ & $0.526$ & $0.047$ & $0.029$ & $0.349$ & $15593$ \\ 
    $0.500$ & $0.7071$ & $450072$ & $0.065$ & $0.482$ & $0.055$ & $0.033$ & $0.365$ & $16272$ \\ 
    $0.475$ & $0.7246$ & $584371$ & $0.084$ & $0.379$ & $0.151$ & $0.037$ & $0.349$ & $20957$ \\ 
    $0.450$ & $0.7416$ & $718286$ & $0.114$ & $0.312$ & $0.194$ & $0.041$ & $0.340$ & $35346$ \\ 
    $0.425$ & $0.7583$ & $975629$ & $0.151$ & $0.229$ & $0.184$ & $0.095$ & $0.341$ & $68338$ \\ 
    $0.400$ & $0.7746$ & $1202753$ & $0.181$ & $0.188$ & $0.209$ & $0.096$ & $0.326$ & $76230$ \\ 
    $0.375$ & $0.7906$ & $1435734$ & $0.210$ & $0.160$ & $0.224$ & $0.093$ & $0.312$ & $77970$ \\ 
    $0.350$ & $0.8062$ & $1739772$ & $0.254$ & $0.133$ & $0.236$ & $0.087$ & $0.291$ & $80100$ \\ 
    $0.325$ & $0.8216$ & $2059217$ & $0.288$ & $0.117$ & $0.239$ & $0.083$ & $0.273$ & $82714$ \\ 
    $0.300$ & $0.8367$ & $2383804$ & $0.316$ & $0.102$ & $0.244$ & $0.080$ & $0.258$ & $84953$ \\ 
    $0.275$ & $0.8515$ & $2728214$ & $0.350$ & $0.090$ & $0.243$ & $0.078$ & $0.239$ & $86151$ \\ 
    $0.250$ & $0.8660$ & $3071192$ & $0.374$ & $0.083$ & $0.240$ & $0.076$ & $0.226$ & $90357$ \\ 
    $0.225$ & $0.8803$ & $3420697$ & $0.396$ & $0.078$ & $0.239$ & $0.074$ & $0.213$ & $94210$ \\ 
    $0.200$ & $0.8944$ & $3807556$ & $0.416$ & $0.076$ & $0.237$ & $0.073$ & $0.199$ & $101358$ \\ 
    $0.175$ & $0.9083$ & $4210208$ & $0.432$ & $0.074$ & $0.234$ & $0.072$ & $0.188$ & $102774$ \\ 
    $0.150$ & $0.9220$ & $4651704$ & $0.446$ & $0.072$ & $0.233$ & $0.073$ & $0.177$ & $103831$ \\ 
    $0.125$ & $0.9354$ & $5049016$ & $0.455$ & $0.069$ & $0.235$ & $0.073$ & $0.167$ & $104227$ \\ \hline 
    
\end{tabular}
\caption{\label{tab4} \small For each fixed values of $bar w$, we
  computed the percentage of proteins, among those belonging to the
  largest component, that come from the five kingdoms.}
 \end{center}
\end{table*}

Starting from approximately $\bar{w}\sim 0.65$ the largest component
begins to expand its size capturing a lot of smaller components.
Furthermore the components which are disconnetted at $\bar{w} \sim
0.675$ and which go to form the giant component at $\bar{w}\sim 0.65$
are samples of many different sizes, from small components to very big
components. This phenomenon becomes more and more evident for lower
values of $\bar{w}$, when the coordination degree distribution of the
giant component follows a power law scaling.  This is evident also
from Figure~\ref{fig6b}, where we plot the distribution of the
coordination degree for the whole set of proteins. The exponent
$\alpha(\bar{w})$ of the power law behavior $f_{\bar w}(z) \sim
z^{-\alpha(\bar w)}$ varies slightly between the regions corresponding
to small values of the coordination degree $z$ and to large values of
$z$. Clearly when a giant component exists, the region with large $z$
is largely determined by the giant component itself. In
Table~\ref{tab3} we report the fitting values of the exponent
$\alpha(\bar{w})$ computed in two regions with small and large values
of $z$. As we decrease the value of $\bar w$, the two fitting values
of $\alpha(\bar w)$ become more and more divergent. In fact, since the
largest component is growing, the tail of the distribution $f_{\bar
w}(z)$ becomes more and more important and assumes a power law
behavior characterized by a different exponent.

A significant fact goes with the rapid size increase of the largest
component. In Table~\ref{tab4} we show, for each $\bar{w}$, the fraction of
different kingdoms and the number of different species which appear in
the largest connected component.  Down to around $\bar{w} = 0.675 $
only proteins coming from viruses belong to the largest component and,
moreover this largest cluster has not yet become giant with respect to
smaller clusters. For $\bar{w} \lesssim 0.675$ the formation of a
giant component begins, and simultaneously all kinds of kingdoms enter
in the species composition of the giant cluster. This is also evident
from Figure~\ref{fig9}b, where we plot the fraction of the number of species
belonging to the largest component.  This ratio increases rapidly
around the same value of $\bar w$. These processes continue for lower
values of $\bar{w}$, with the giant component including more and more
proteins belonging to many different species, and the ratio for each
kingdom tends to become the same as that of the whole database.
Furthermore around $\bar w \simeq 0.475$ there is a very sharp
increase both in the dimension of the giant component and especially
in the number of species present in it, as it is evident from Figures
~\ref{fig9}a and ~\ref{fig9}b.

The processes just described may indicate the presence of a phase
transition: we have two different phases, one for large values of
$\bar w$, characterized by the presence of clusters with similar
dimensions and with the largest one composed especially of viruses,
and the second phase characterized by the presence of a giant
component composed of different species alongside other small little
clusters. We note however that the phase transition is not sharp, but
the changes in the dimension and composition of the largest component
are spread in a range $ 0.475 < \bar w <0.675$. We also note that the
plot in Figure~\ref{fig9}b has a very rapid increase for $w \sim 0.475$.

\begin{table*}[!htb]
  \begin{center} 
  \begin{tabular}{|c|c|c|c|c|c|c|c|c|c|c|c|c|c|c|c|}  \hline 
	
    $\bar{w}$ & $0.95$ & $0.90$ & $0.85$ & $0.80$ &
    $0.75$ & $0.70$ & $0.65$ & $0.60$ & $0.55$ &
    $0.50$ & $0.45$ & $0.35$ & $0.25$ & $0.15$ \\ \hline 

    bacteria & $9.6$ & $12.2$ & $14.2$ & $17.2$ & $21.9$ & $22.6$
    & $23.6$ & $23.8$ & $23.9$ & $25.1$ & $25.8$ & $29.0$ & $35.6$
    & $57.0$ \\
    
    viruses & $32.7$ & $31.4$ & $24.3$ & $17.6$ & $11.4$ & $7.4$ &
    $5.2$ & $3.8$ & $2.9$ & $2.7$ & $2.4$ & $2.7$ & $4.2$ & $7.5$
    \\
    
    plants & $9.3$ & $10.8$ & $11.4$ & $9.4$ & $8.3$ & $7.3$ &
    $7.6$ & $7.8$ & $7.7$ & $7.5$ & $7.5$ & $6.2$ & $4.0$ & $0.0$
    \\
    
    invertebrates & $11.6$ & $8.9$ & $7.4$ & $5.8$ & $3.6$ & $3.2$
    & $2.5$ & $2.0$ & $1.6$ & $1.5$ & $1.2$ & $1.4$ & $1.3$ &
    $1.1$ \\
    
    vertebrates & $22.9$ & $23.0$ & $25.4$ & $25.7$ & $25.6$ &
    $25.9$ & $23.6$ & $20.0$ & $17.1$ & $13.0$ & $10.2$ & $5.2$ &
    $2.8$ & $1.1$ \\
    
    bac-vir & $2.7$ & $2.2$ & $2.1$ & $2.1$ & $1.6$ & $1.6$ &
    $1.4$ & $1.0$ & $1.0$ & $1.1$ & $1.0$ & $1.7$ & $2.4$ & $3.2$
    \\
    
    bac-pla & $1.6$ & $1.8$ & $2.8$ & $2.9$ & $3.5$ & $4.5$ &
    $5.9$ & $7.0$ & $8.5$ & $8.9$ & $9.1$ & $10.8$ & $11.3$ &
    $18.3$ \\
    
    bac-inv & $0.5$ & $0.4$ & $0.7$ & $0.7$ & $0.8$ & $0.9$ &
    $1.3$ & $1.7$ & $2.1$ & $2.1$ & $2.0$ & $2.6$ & $3.0$ & $1.1$
    \\
    
    bac-ver & $1.8$ & $2.0$ & $2.4$ & $2.3$ & $1.9$ & $1.9$ &
    $1.8$ & $1.6$ & $1.5$ & $1.5$ & $1.3$ & $1.1$ & $1.1$ & $1.1$
    \\
    
    vir-pla & $0.2$ & $0.1$ & $0.2$ & $0.4$ & $0.3$ & $0.4$ &
    $0.3$ & $0.3$ & $0.2$ & $0.2$ & $0.2$ & $0.2$ & $0.5$ & $0.0$
    \\
    
    vir-inv & $0.0$ & $0.0$ & $0.0$ & $0.0$ & $0.0$ & $0.0$ &
    $0.0$ & $0.0$ & $0.0$ & $0.0$ & $0.0$ & $0.0$ & $0.0$ & $0.0$
    \\
    
    vir-ver & $0.2$ & $0.5$ & $0.7$ & $0.8$ & $0.9$ & $0.7$ &
    $0.6$ & $0.4$ & $0.3$ & $0.2$ & $0.1$ & $0.2$ & $0.1$ & $0.0$
    \\
    
    pla-inv & $0.9$ & $0.0$ & $0.0$ & $0.0$ & $0.0$ & $0.2$ &
    $0.1$ & $0.1$ & $0.1$ & $0.2$ & $0.3$ & $0.2$ & $0.5$ & $0.0$
    \\
    
    pla-ver & $0.5$ & $0.9$ & $0.8$ & $1.1$ & $1.3$ & $1.0$ &
    $1.1$ & $1.2$ & $1.2$ & $1.0$ & $0.9$ & $1.3$ & $1.7$ & $1.1$
    \\
    
    inv-ver & $0.5$ & $1.1$ & $2.6$ & $4.5$ & $7.0$ & $8.4$ &
    $9.2$ & $10.3$ & $10.9$ & $11.2$ & $11.0$ & $9.0$ & $5.5$ &
    $0.0$ \\
    
    bac-vir-pla & $0.0$ & $0.4$ & $0.3$ & $0.5$ & $0.3$ & $0.3$ &
    $0.4$ & $0.2$ & $0.2$ & $0.2$ & $0.4$ & $0.4$ & $0.7$ & $1.1$
    \\
    
    bac-vir-inv & $0.0$ & $0.0$ & $0.0$ & $0.0$ & $0.0$ & $0.0$ &
    $0.0$ & $0.0$ & $0.1$ & $0.0$ & $0.1$ & $0.1$ & $0.3$ & $1.1$
    \\
    
    bac-vir-ver & $0.2$ & $0.1$ & $0.0$ & $0.0$ & $0.0$ & $0.1$ &
    $0.1$ & $0.1$ & $0.1$ & $0.1$ & $0.2$ & $0.2$ & $0.2$ & $0.0$
    \\
    
    bac-pla-inv & $0.0$ & $0.0$ & $0.1$ & $0.2$ & $0.5$ & $0.6$ &
    $0.8$ & $0.9$ & $1.3$ & $2.0$ & $2.3$ & $2.4$ & $3.1$ & $1.1$
    \\
    
    bac-pla-ver & $0.0$ & $0.1$ & $0.0$ & $0.0$ & $0.1$ & $0.3$ &
    $0.6$ & $0.6$ & $0.9$ & $1.0$ & $1.3$ & $1.7$ & $1.4$ & $0.0$
    \\
    
    bac-inv-ver & $0.0$ & $0.0$ & $0.1$ & $0.3$ & $0.4$ & $0.4$ &
    $0.4$ & $0.9$ & $0.8$ & $0.9$ & $0.9$ & $1.0$ & $0.8$ & $1.1$
    \\
    
    vir-pla-inv & $0.0$ & $0.0$ & $0.0$ & $0.0$ & $0.0$ & $0.0$ &
    $0.0$ & $0.0$ & $0.0$ & $0.0$ & $0.0$ & $0.0$ & $0.0$ & $0.0$
    \\
    
    vir-pla-ver & $0.0$ & $0.0$ & $0.0$ & $0.0$ & $0.1$ & $0.1$ &
    $0.1$ & $0.0$ & $0.0$ & $0.0$ & $0.0$ & $0.0$ & $0.0$ & $0.0$
    \\
    
    vir-inv-ver & $0.0$ & $0.0$ & $0.1$ & $0.2$ & $0.1$ & $0.2$ &
    $0.3$ & $0.2$ & $0.2$ & $0.2$ & $0.2$ & $0.1$ & $0.1$ & $0.0$
    \\
    
    pla-inv-ver & $0.9$ & $1.4$ & $1.8$ & $5.5$ & $7.3$ & $8.4$ &
    $9.4$ & $11.0$ & $11.3$ & $12.0$ & $12.4$ & $13.4$ & $11.7$ &
    $0.0$ \\
    
    bac-vir-pla-inv & $0.0$ & $0.0$ & $0.0$ & $0.0$ & $0.0$ &
    $0.0$ & $0.0$ & $0.1$ & $0.1$ & $0.1$ & $0.1$ & $0.2$ & $0.0$
    & $0.0$ \\
    
    bac-vir-pla-ver & $0.0$ & $0.0$ & $0.1$ & $0.1$ & $0.1$ &
    $0.1$ & $0.2$ & $0.1$ & $0.1$ & $0.1$ & $0.1$ & $0.1$ & $0.1$
    & $0.0$ \\
    
    bac-vir-inv-ver & $0.0$ & $0.0$ & $0.0$ & $0.0$ & $0.0$ &
    $0.0$ & $0.0$ & $0.0$ & $0.0$ & $0.0$ & $0.0$ & $0.1$ & $0.1$
    & $0.0$ \\
    
    bac-pla-inv-ver & $0.2$ & $0.1$ & $0.4$ & $0.7$ & $1.0$ &
    $2.1$ & $2.5$ & $3.8$ & $5.1$ & $6.4$ & $8.0$ & $7.6$ & $6.7$
    & $0.0$ \\
    
    vir-pla-inv-ver & $0.0$ & $0.1$ & $0.0$ & $0.0$ & $0.1$ &
    $0.1$ & $0.2$ & $0.3$ & $0.3$ & $0.3$ & $0.2$ & $0.2$ & $0.1$
    & $1.1$ \\
    
    bac-vir-pla-inv-ver & $0.0$ & $0.0$ & $0.1$ & $0.1$ & $0.1$ &
    $0.2$ & $0.2$ & $0.1$ & $0.2$ & $0.3$ & $0.5$ & $0.7$ & $0.4$
    & $1.1$ \\ \hline
    
\end{tabular}
\caption{\label{tab5} \small Spread of species in connected
  components. Each value indicates the percentage of clusters,
  calculated on clusters having size greater than $90$, composed by
  proteins coming from only one kingdom, only from a pair of kingdoms,
  etc., up to the percentage of clusters composed by proteins of all
  kingdoms.}
 \end{center}
\end{table*}

In Table~\ref{tab5}, for each $\bar{w}$, it can be seen how different
kingdoms are distributed in connected components. In particular we
count the number of components, whose size is greater than $90$ and
record the percentage of clusters whose proteins come from species of
only one kingdom, only from a pair of kingdoms, etc., up to the
percentage of connected components which contain proteins of all
kingdoms. For high values of $\bar{w}$ the majority of clusters are
made up of proteins belonging to only one kingdom, in particular the
kingdom of viruses; clusters with proteins of different kingdoms are
very scarce. As expected, as $\bar{w}$ decreases, the percentage of
clusters belonging to only one kingdom decreases in favor of clusters
of mixed kingdom composition.

It is interesting to note that the virus kingdom has a very low
tendency to cluster with the other kingdoms, in particular with plants
and animalia. Furthermore, for no values of $\bar{w}$ do we see the
formation of components (of size greater than $90$) with proteins
coming from viruses and invertebrates, and from viruses, plants and
invertebrates. Virus proteins cluster mainly with bacterial proteins.
In addition we observe that bacterial proteins cluster mainly with
plant proteins and vice versa. Moreover, although plant proteins
cluster infrequently with invertebrates and with vertebrates, there
are many more clusters consisting simultaneously of plant,
invertebrate and vertebrate proteins. Finally we note that at the
lowest value of $\bar{w}$, the majority of components which are not
included in the giant component are clusters consisting of bacterial
proteins, of bacterial and plant proteins and of virus proteins.

\section{Analysis of the proteins that connect clusters}

\begin{figure}[!htb]
  \begin{center}
    \vspace{-0.4cm}
    \subfigure[]{\label{fig10a} 
      \includegraphics[width=0.48\textwidth]{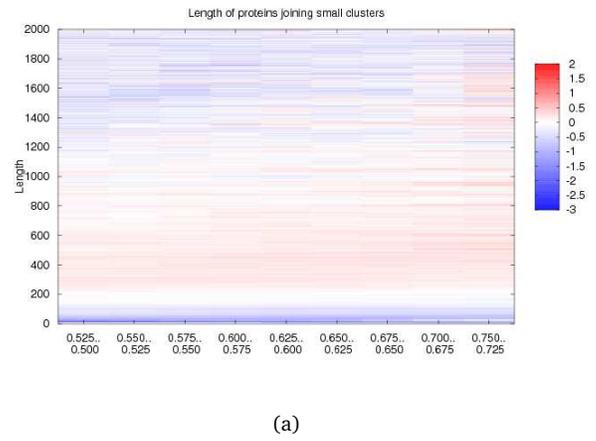}
    } \vspace{-0.4cm}
    \subfigure[]{\label{fig10b}
      \includegraphics[width=0.48\textwidth]{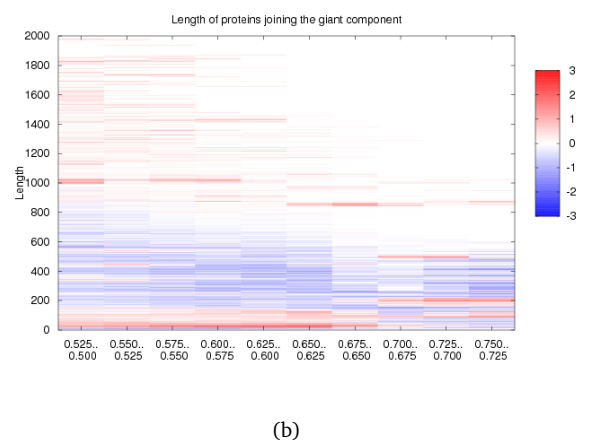}
    }
    \caption{{\small Length representation of (a) proteins joining
        generic clusters and of (b) proteins joining the largest
        cluster. The red color encodes overrepresented lengths; the
        blue color indicates underrepresented lengths.}}
  \end{center}
\end{figure}

\begin{figure}[!htb]
  \begin{center}
    \vspace{-0.4cm}
    \subfigure[\label{fig11a}]{
     \includegraphics[width=0.48\textwidth]{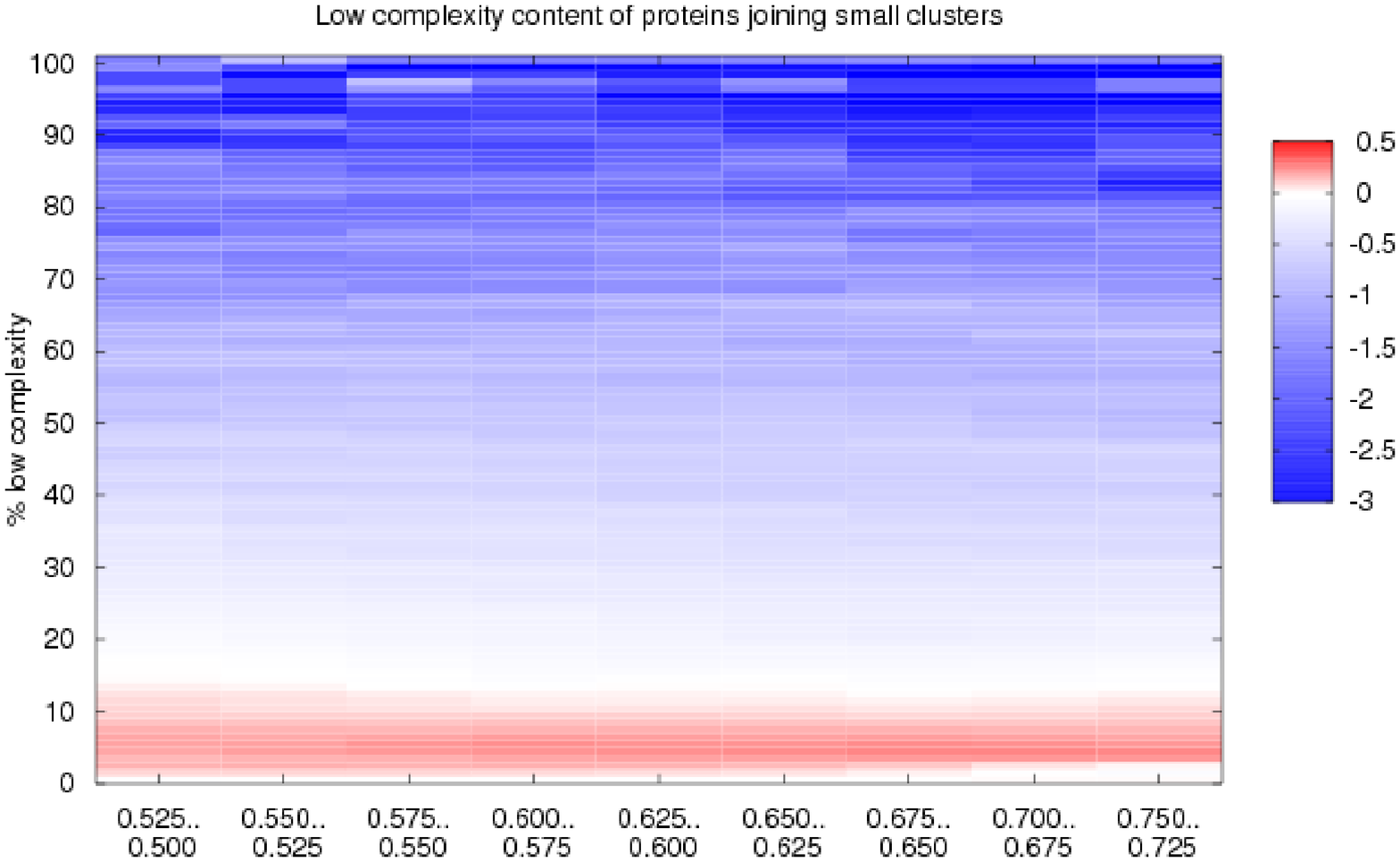}
    }\vspace{-0.4cm}
    \subfigure[\label{fig11b}]{
      \includegraphics[width=0.48\textwidth]{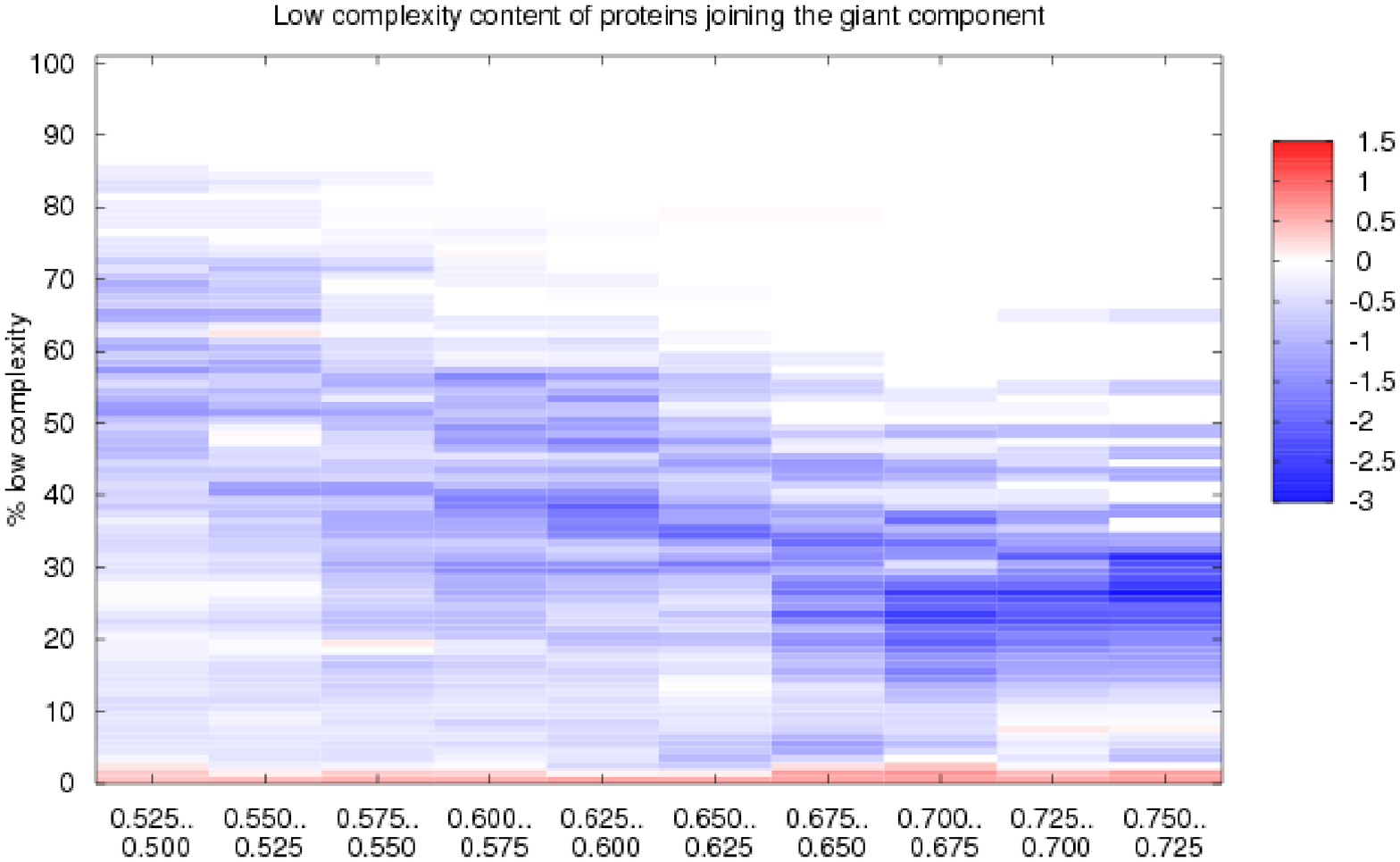}
    }
  \caption{{\small Representation of the low complexity
      content of (a) proteins joining generic clusters and of (b)
      proteins joining the largest cluster. The red color encodes
      overrepresented values; the blue color indicates
      underrepresented values.  }}
  \end{center}
\end{figure}

\begin{figure}[!htb]
  \begin{center}
    \vspace{-0.4cm}
    \subfigure[\label{fig12a}]{
     \includegraphics[width=0.48\textwidth]{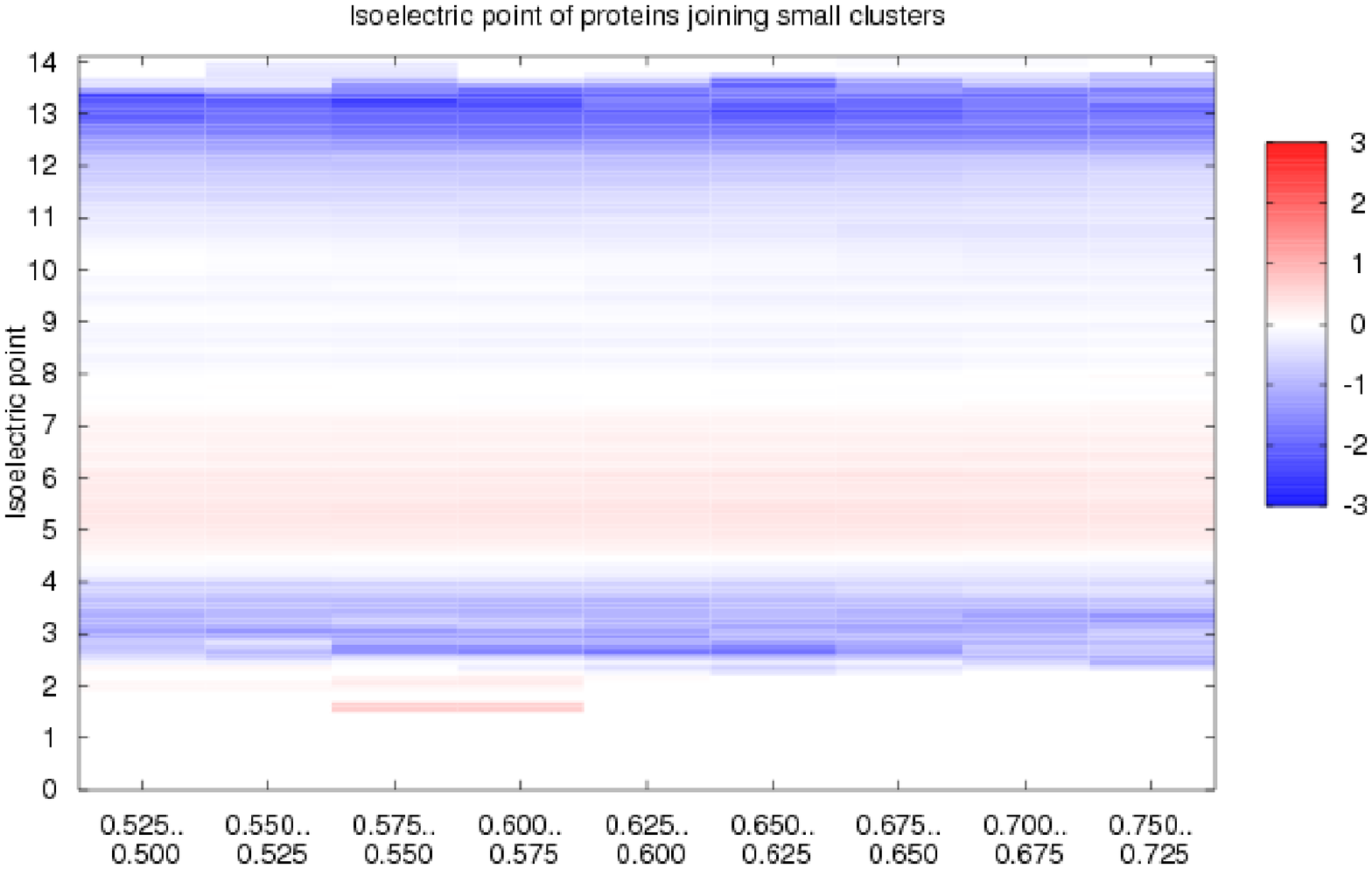}
    }\vspace{-0.4cm}
    \subfigure[\label{fig12b}]{
      \includegraphics[width=0.48\textwidth]{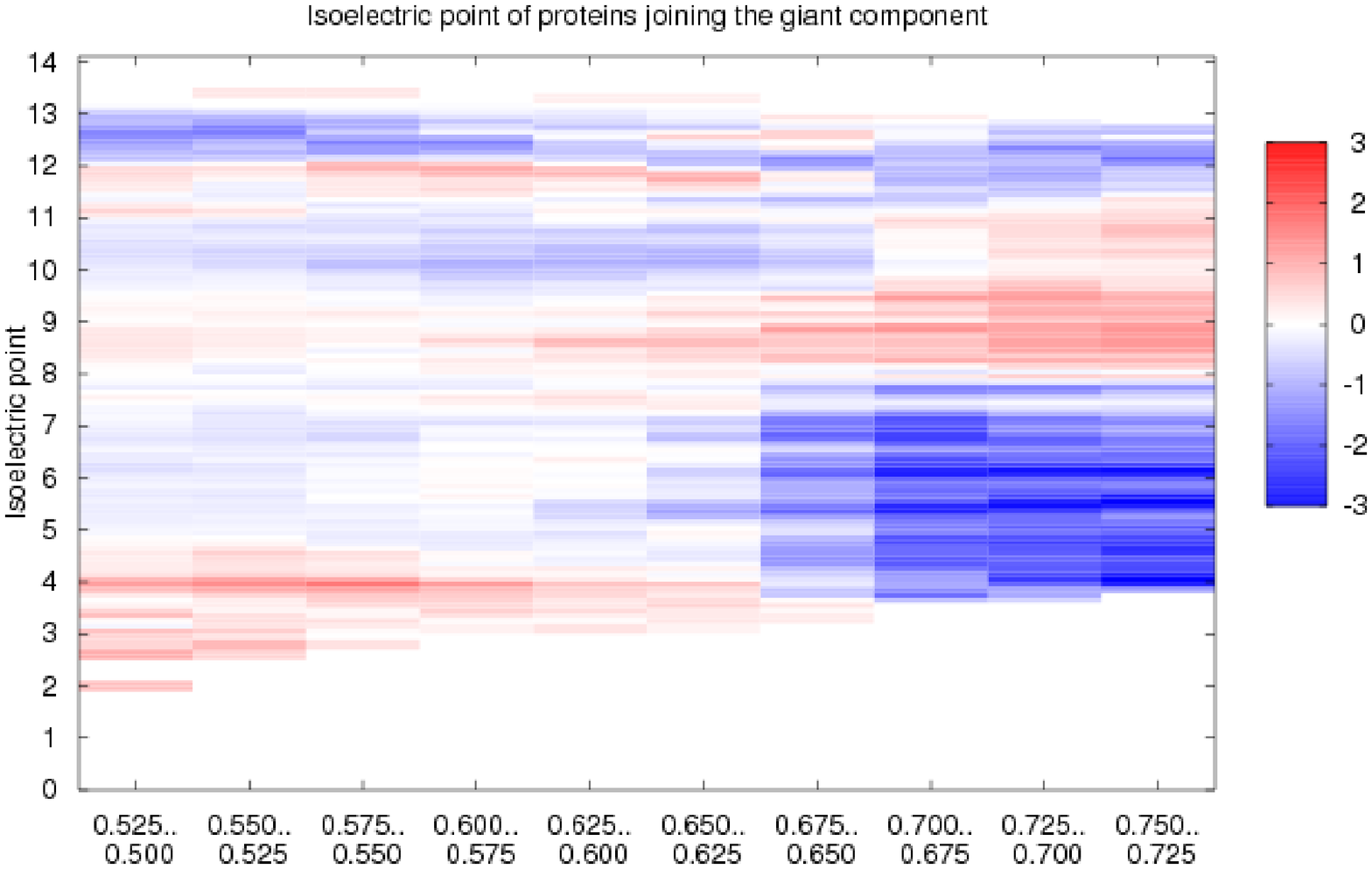}
    }
\caption{{\small Representation of the isoelectric
      points of (a) proteins joining generic clusters and of (b)
      proteins joining the largest cluster.  The red color encodes
      overrepresented values; the blue color indicates
      underrepresented values.}}
  \end{center}
\end{figure}

\begin{figure}[!htb]
  \begin{center}
    \vspace{-0.4cm}
    \subfigure[\label{fig13a}]{
      \includegraphics[width=0.48\textwidth]{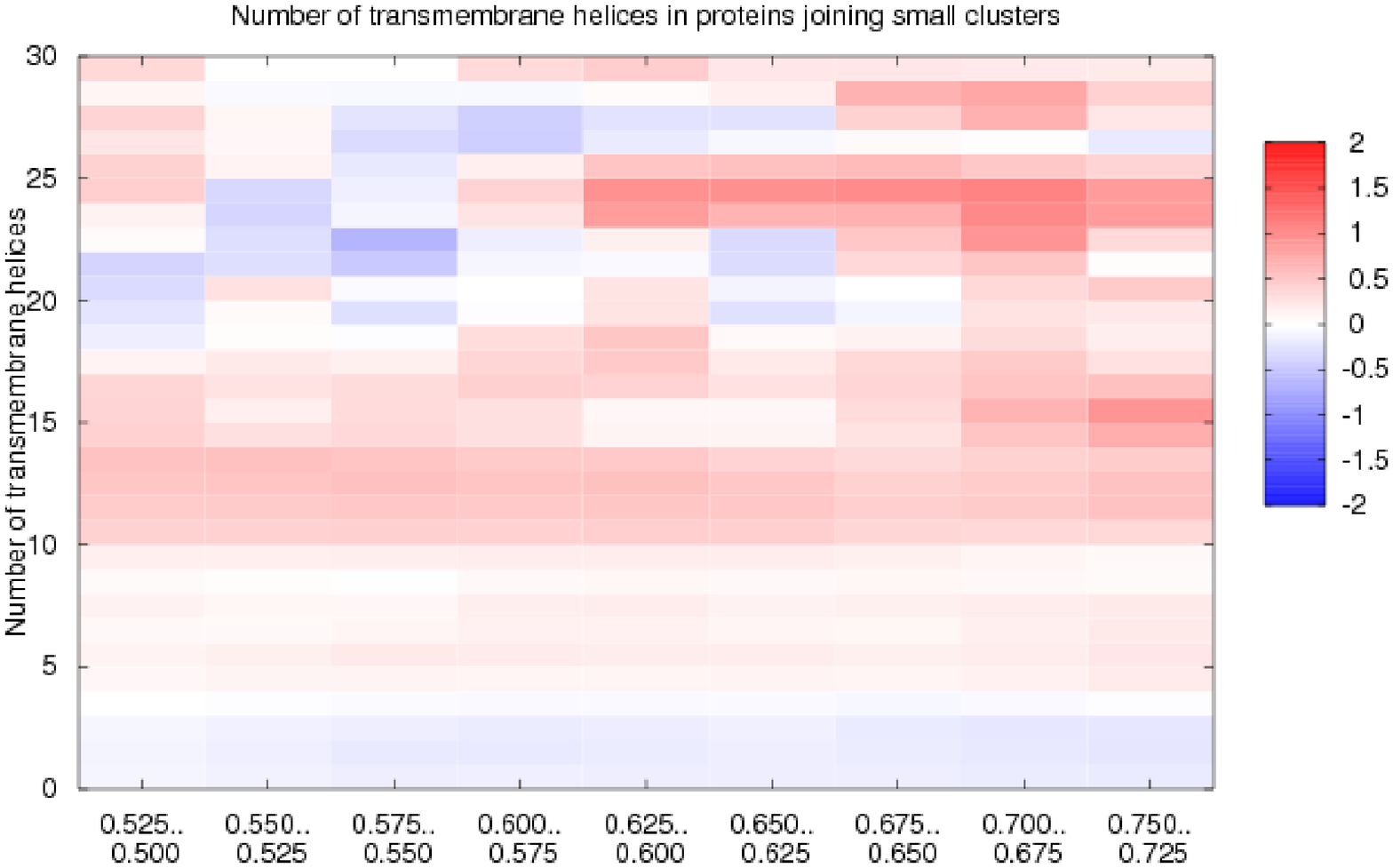}
    }\vspace{-0.4cm}
    \subfigure[\label{fig13b}]{
      \includegraphics[width=0.48\textwidth]{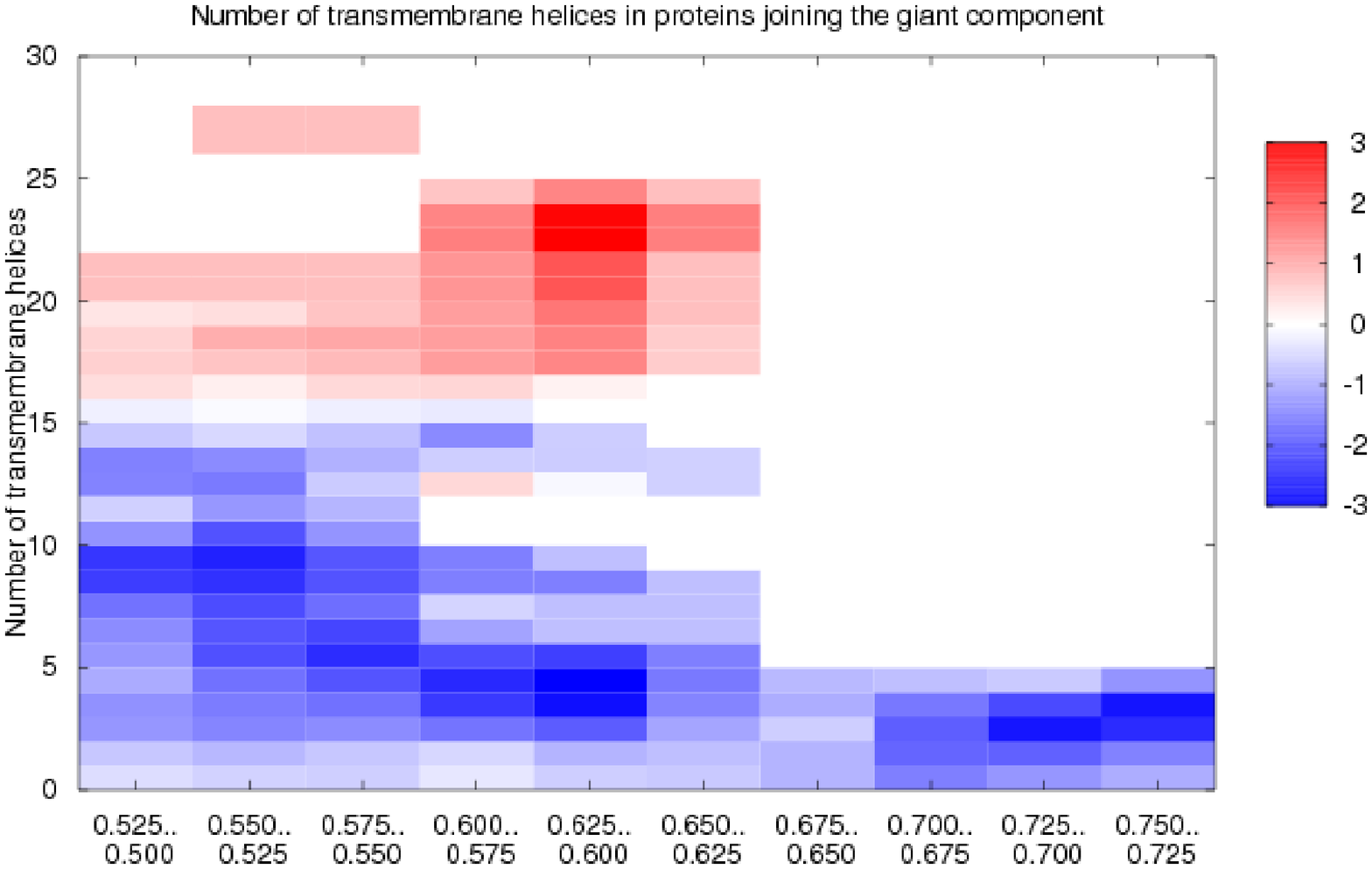}
    }
    \caption{{\small Representation of the predicted number
	of transmembrane helices of (a) proteins joining generic
	clusters and of (b) proteins joining the largest cluster. The
	red color encodes overrepresented values; the blue color
	indicates underrepresented values.}}
  \end{center}
\end{figure}

\begin{figure}[!htb]
  \begin{center}
    \vspace{-0.4cm}
    \subfigure[\label{fig14a}]{
     \includegraphics[width=0.48\textwidth]{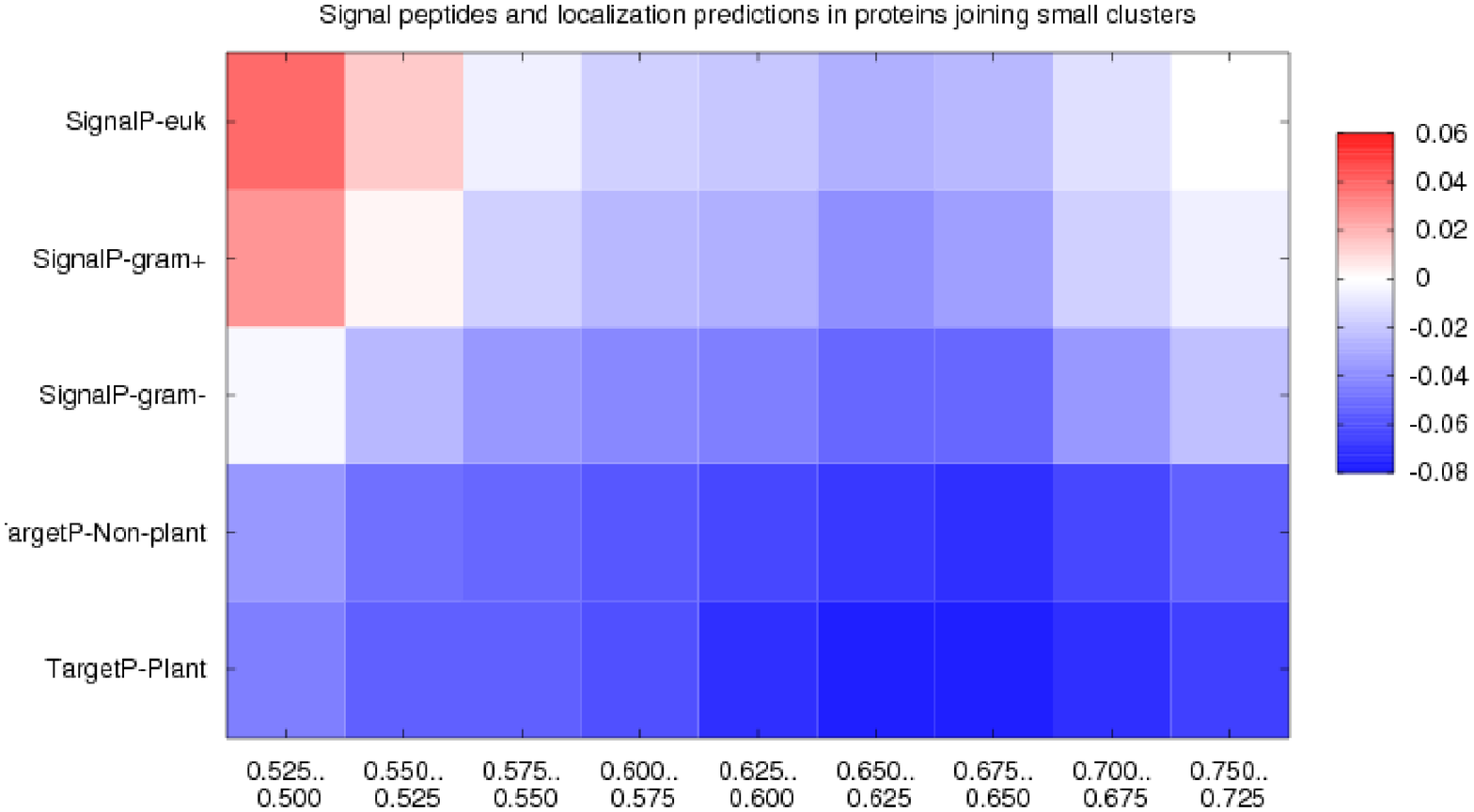}
    }\vspace{-0.4cm}
    \subfigure[\label{fig14b}]{
      \includegraphics[width=0.48\textwidth]{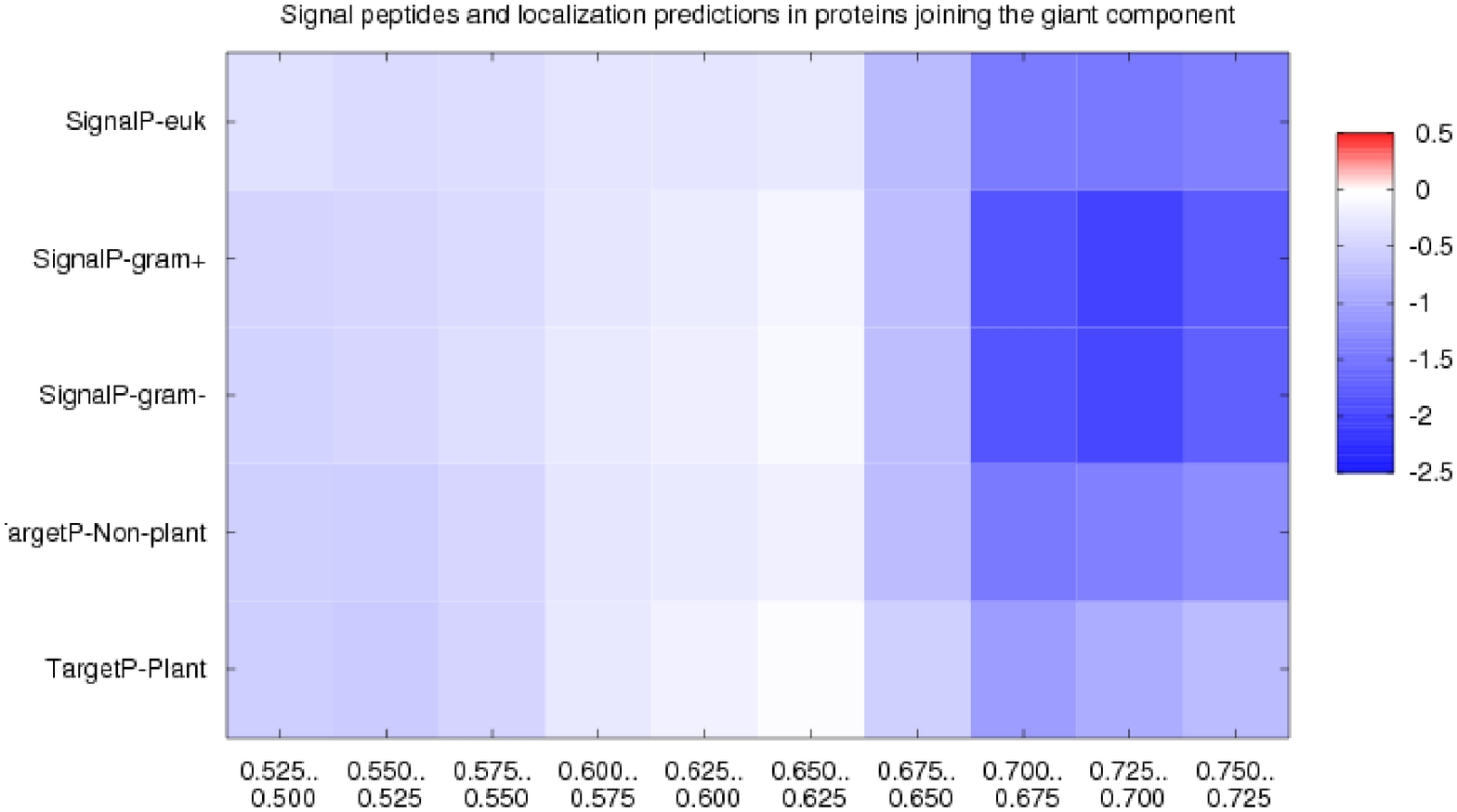}
    }
\caption{{\small Representation of the predicted signal peptides and
      protein localization signals of (a) proteins joining generic
      clusters and of (b) proteins joining the largest cluster. The
      red color encodes overrepresented values; the blue color
      indicates underrepresented values. }}
  \end{center}
\end{figure}

\begin{figure}[!htb]
  \begin{center}
    \vspace{-0.4cm}
    \subfigure[\label{fig15a}]{
     \includegraphics[width=0.48\textwidth]{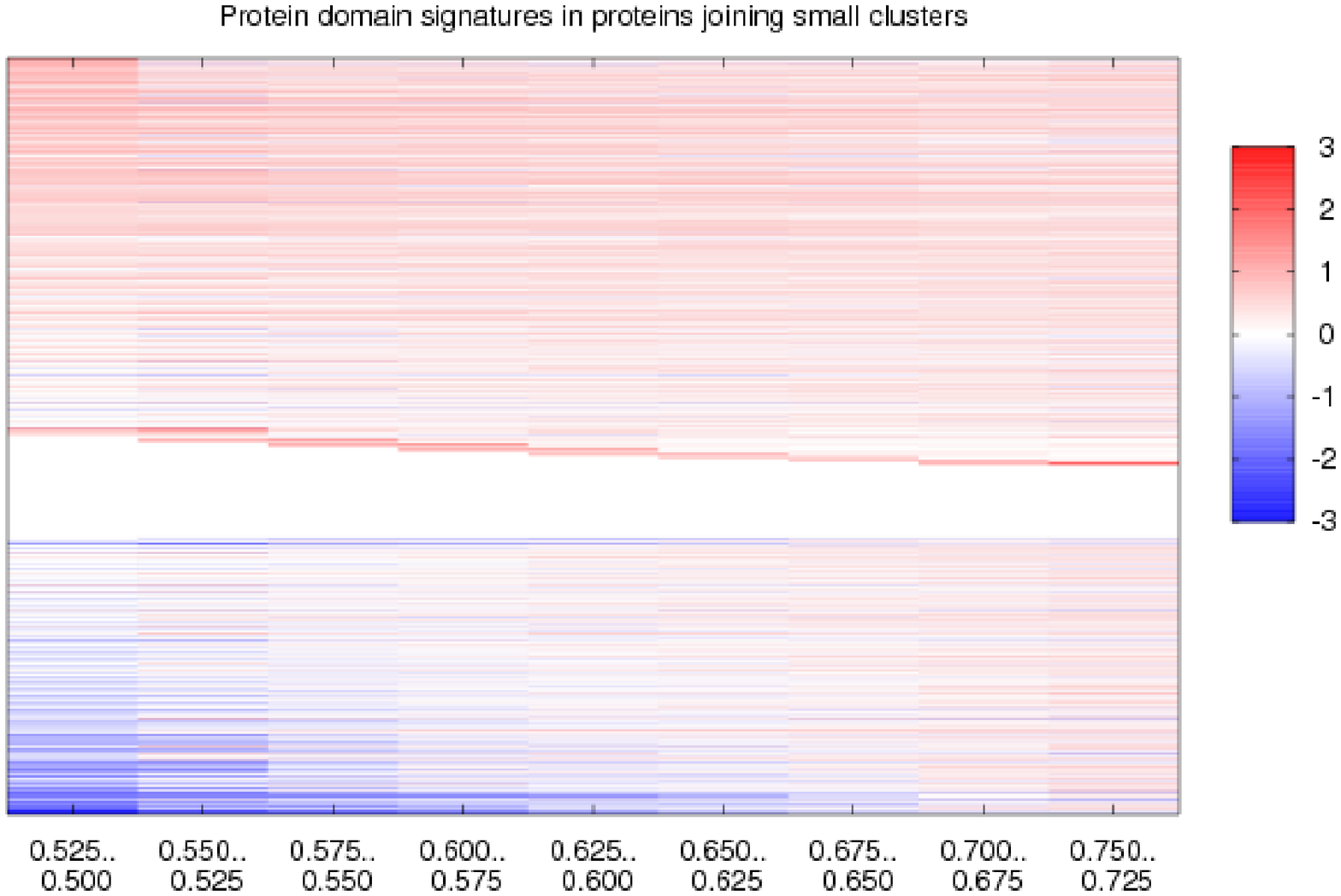}
    }\vspace{-0.4cm}
    \subfigure[\label{fig15b}]{
      \includegraphics[width=0.48\textwidth]{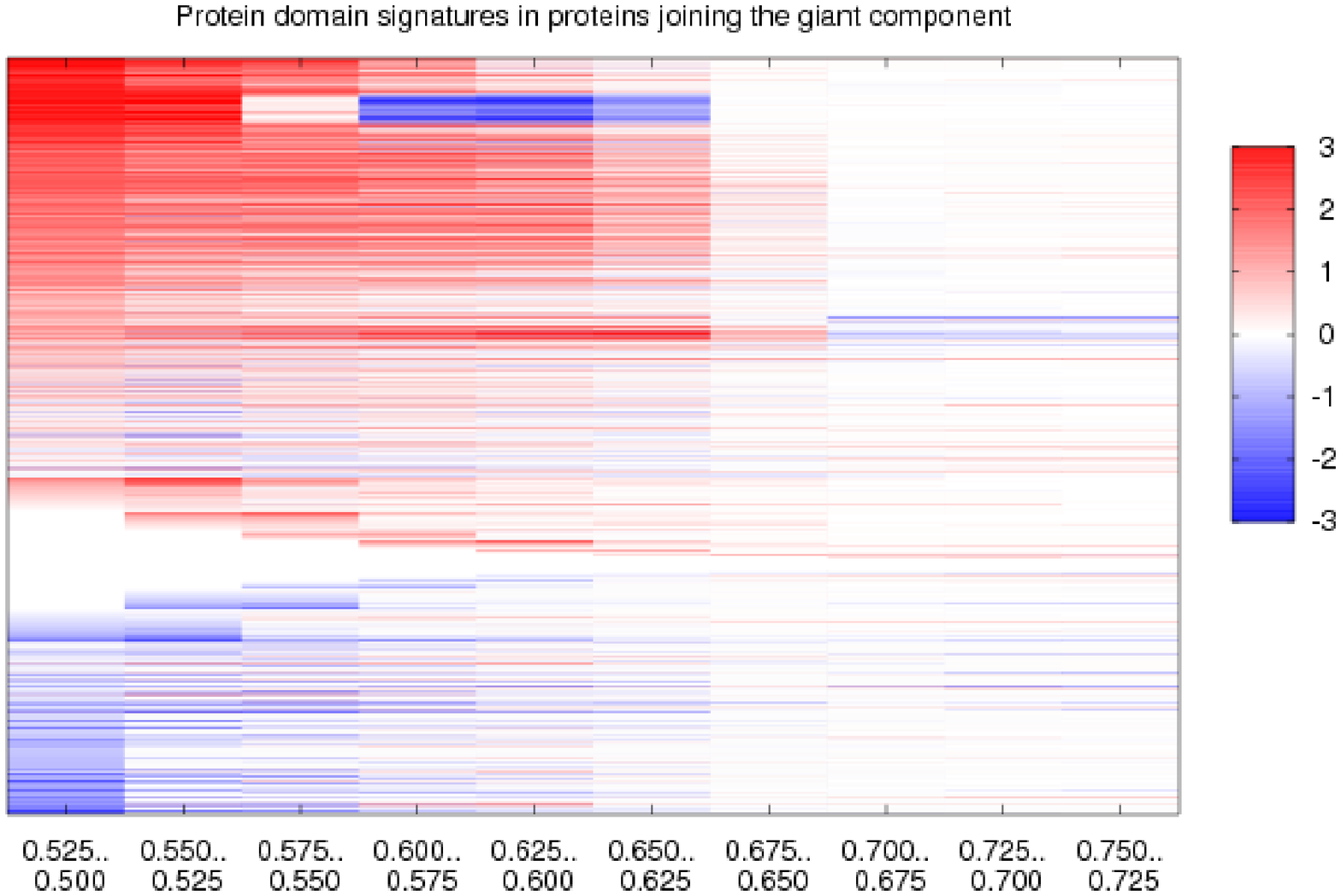}
    }
\caption{{\small Representation of the predicted protein domains of
      (a) proteins joining generic clusters and of (b) joining the
      largest cluster.  Each line in the graph denotes a certain
      domain. The red color encodes overrepresented values; the blue
      color indicates underrepresented values.}}
  \end{center}
\end{figure}

Protein pairs that connect clusters in the different weight intervals
are of special interest as they harbor the most conserved sequence
regions that are shared by the interconnected clusters. We want to
know if certain sequence features and protein domains are enriched in
these proteins compared to the complete proteome.  Therefore we have
calculated for all protein contained in SIMAP some sequence features:
\e{length}, \e{isoelectric point} (using the EMBOSS sequence analysis
package \cite{emboss}), \e{low complexity content} (using the program
seg \cite{segprog}) and the number of \e{predicted transmembrane
segments} (using the program TMHMM \cite{tmhmmprog}).  Additionally,
in order to derive functional information for all proteins, we have
predicted \e{signal peptides} (using SignalP 3.0 \cite{signalPprog}),
\e{localization signals} (using TargetP 1.1\cite{targetPprog}) and
\e{protein domains} (using the databases PFAM, TIGRFAM, PANTHER,
SUPERFAMILY, SMART and PIRSF from InterPro 12.1 \cite{pddb}) for all
SIMAP proteins.

For all weight intervals we have counted the feature occurrence in the
proteins that connect clusters; these proteins are all pairs of
sequences which belong to different clusters in the graph built at
$\bar{w}_1$ and belonging to the same cluster in the graph built at
$\bar{w}_2$, where $\bar{w}_2<\bar{w}_1$ are two consecutive values of
the weight $\bar{w}$. We have also distinguished between two disjoint
sets of these proteins: proteins linking the clusters that will form
the largest cluster in the graph built at $\bar{w}_2$ and proteins
linking the other generic clusters.

The enrichment ($e$) of features was calculated as ratio of the number
of features found ($k$) and the number of features expected ($k_E$):
$e = k/k_E$. The number of features expected was calculated by: $k_E =
K\,n/V$, where $n$ is the number of proteins of interest (e.g.
connecting clusters in a given weight interval), $K$ denotes the
number of proteins used for clustering having the given feature and
$V$ corresponds to the number of proteins used for clustering.

\subsection{Results}

Proteins joining clusters outside the largest cluster show an
over-representation of lengths around 400aa (Figure~\ref{fig10a}),
contain overrepresented proteins of small low complexity content
(Figure~\ref{fig11a}), are often neutral or weakly acidic
(Figure~\ref{fig12a}) and contain more transmembrane proteins than
expected (Figure~\ref{fig13a}).  Proteins joining clusters in the
giant component are characterized by short and very long lengths
(Figure~\ref{fig10b}), reduced low complexity content
(Figure~\ref{fig11b}), acidic and alkaline proteins, dependent on the
weight interval (Figure~\ref{fig12b}) and a high number of
transmembrane domains in the lower weight intervals
(Figure~\ref{fig13b}).  Signal peptides were found overrepresented in
proteins joining clusters outside the largest component at the lower
weight intervals; at higher weight intervals and in proteins joining
clusters in the largest component they were found underrepresented, as
were localization signals in all proteins joining clusters
(Figure~\ref{fig14a} and Figure~\ref{fig14b}).  For all considered
weight intervals we could find interval-specific overrepresented and
underrepresented protein domains (Figure~\ref{fig15a} and
\ref{fig15b}). Remarkably these domains are not only specific for a
certain weight interval, but also different for proteins joining
clusters outside the largest component and proteins joining clusters
in the largest component (See Table~\ref{tab6}).

\subsection{Discussion}

All of the analyzed sequence features indicate that proteins that join
clusters at a certain weight interval are not distributed equally over
the complete protein space. For all of the features we could find
specific under- and over-representation. Proteins joining clusters
outside the largest component and proteins joining clusters in the
largest component are different with respect to almost all considered
features, which indicates that the largest component contains proteins
that are different from those contained in other large clusters. These
findings are complemented by the observation of specific over- and
underrepresented functional domains in the proteins connecting
clusters at certain weight intervals. Thus we conclude that for each
weight interval a small number of protein families is responsible for
cluster interconnections.

\section{Conclusions}

We investigated the local e global properties of the sequence
similarity space formed by all proteins in the SIMAP database, which
contains more than $5.5$ millon amino acid sequences. We represented
this space as a graph whose vertices are proteins and the edges are
weighted to reflect the similarity between the corresponding pairs of
sequences (high weight, high similarity). The choice of this weight
formula (\ref{eq:weight}) came from the necessity to compare the
similarity score between pairs of sequences that could have different
lengths. The SW score was therefore modified by means of the
self-score geometric mean which contains the length information of the
two aligned sequences.

Then, keeping only edges with $w \geq \bar w$, we built a collection
of graphs by varing $\bar w$. From the analysis of the connected
components we found that these graphs do not belong to the class of
random graphs, whereas they are characterized by a power law behaviour
both in the size cluster distribution and in the coordination degree
distribution and for each fixed $\bar w$ these two distributions are
strongly related to each other.

With the variation of $\bar w$, we found interesting changes in the
global organization of the protein homology networks: we observed two
different phases, one for large values of $\bar w $, characterized by
the presence of clusters with similar dimensions, each composed
essentially by proteins belonging to only one kingdom and with the
largest one composed especially by viruses, and the second phase, for
lower values of $\bar w$, characterized by the presence of a giant
component composed by different species and other very little
clusters.

In the end we investigated sequence features and functional
informations of protein pairs that are responsible of the connection
of clusters in the different intervals of $\bar w$, since they harbor
the most conserved sequence regions that are shared by the
interconnected clusters. We found that proteins joining clusters
outside the largest component and proteins joining clusters in the
largest component are different with respect to almost all considered
features, which indicates that the largest component contains proteins
that are different from those contained in other large
clusters. Indeed we found an overrepresentation of a small set of
domains which shows that a small number of protein families is
responsible for cluster interconnections.

The analysis we performed gives a first view of the global
organization of the greatest protein homology network ever been built
before. It is the first step and the starting point to answer to other
global or local interesting questions which could confirm that the
protein homology network is structured with respect to functional and
evolutionary properties.

\section{Acknowledgements}
The authors thanks Claudio Destri, Roland Arnold and Mattia Pelizzola
for useful discussions, Michele Caselle for encouraging our
collaboration and Patrick Tischler, Jan Krumsiek and Benedikt
Wachinger for providing the software for protein feature calculation.

\newpage 
\begin{table*}[!htb] 
  \begin{center}
      \begin{tabular}{|c|cc|cc|}  \hline 
    
    $\bar{w_1} \to \bar{w_2}$ & $e$ & \hspace{-0pt} Proteins joining
    generic clusters & $e$ & \hspace{-0pt} Proteins joining the largest cluster \\
    \hline
    
    & $0.02$	& PF00598 Flu\_M1          & $0.93$	& PF00078 RVT\_1 \\
    & $0.03$	& PF00522 VPR              & $1.08$	& PF00075 RnaseH \\
    & $0.03$	& PF00540 Gag\_p17         & $1.44$	& PF06815 RVT\_connect \\
    & $0.03$	& PF00951 Arteri\_Gl       & $1.46$	& PF07075 DUF1343 \\
    & $0.03$	& PF00971 EIAV\_GP90       & $2.19$	& PF00665 rve \\
    $0.750$ $\to$ $0.725$ & & & & \\
    & $9.40$	& PF02916 DNA\_PPF         & $15.41$	& PF00607 Gag\_p24 \\
    & $11.09$	& PF07095 IgaA             & $18.79$	   & PF00517 GP41 \\ 
    & $11.25$	& PF08272 Topo\_Zn\_Ribbon & $18.91$	& PF02022 Integrase\_Zn \\
    & $11.83$	& PF06899 WzyE             & $27.07$	& PF00540 Gag\_p17 \\
    & $12.46$	& PF06788 UPF0257          & $137.49$	& PF00516 GP120 \\ \hline
    
    & &                                & $0.88$	& PF00078 RVT\_1 \\
    & &                                & $1.16$	& PF00077 RVP \\
    & &                                & $1.91$	& PF06817 RVT\_thumb \\
    & &                                & $3.68$	& PF00075 RnaseH \\
    & &                                & $3.77$	& PF00665 rve \\
    $0.725$ $\to$ $0.700$ & & & & \\ 
    & &                                & $37.19$	& PF00186 DHFR\_1 \\
    & &                                & $80.26$	& PF00098 zf-CCHC \\
    & &                                & $129.77$	& PF00516 GP120 \\
    & &                                & $139.92$	& PF00607 Gag\_p24 \\
    & &                                & $145.50$	& PF00540 Gag\_p17 \\ \hline

    & $0.01$	& PF00516 GP120            & $0.12$	& PF00098 zf-CCHC \\
    & $0.01$	& PF00522 VPR              & $0.15$	& PF00271 Helicase\_C \\
    & $0.01$	& PF00602 Flu\_PB1         & $0.22$	& PF00078 RVT\_1 \\
    & $0.01$	& PF00603 Flu\_PA          & $1.02$	& PF01560 HCV\_NS1 \\
    & $0.01$	& PF01539 HCV\_env         & $1.16$	& PF06817 RVT\_thumb \\
    $0.700$ $\to$ $0.675$ & & & & \\
    & $10.14$	& PF08435 Calici\_coat\_C  & $15.62$	& PF02907 Peptidase\_S29 \\
    & $10.22$	& PF03296 Pox\_polyA\_pol  & $19.47$	& PF00517 GP41 \\
    & $12.94$	& PF05733 Tenui\_N         & $57.66$	& PF00516 GP120 \\
    & $12.98$	& PF03805 CLAG             & $74.03$	& PF00077 RVP \\
    & $13.68$	& PF00897 Orbi\_VP7        & $98.38$	& PF02348 CTP\_transf\_3 \\ \hline
    
    & $0.01$	& PF00064 Neur             & $0.10$	   & PF00078 RVT\_1 \\
    & $0.01$	& PF00469 F-protein        & $0.13$	& PF00077 RVP \\
    & $0.01$	& PF00506 Flu\_NP          & $0.18$	& PF00560 LRR\_1 \\
    & $0.01$	& PF00516 GP120            & $0.18$	& PF00607 Gag\_p24 \\
    & $0.01$	& PF00540 Gag\_p17         & $0.30$	& PF00665 rve \\
    $0.675$ $\to$ $0.650$ & & & & \\
    & $11.63$	& PF04310 MukB             & $151.92$	& PF02959 Tax \\
    & $12.71$	& PF07108 PipA             & $168.64$	& PF00758 EPO\_TPO \\
    & $13.48$	& PF07429 Fuc4NAc\_transf  & $431.37$	& PF08300 HCV\_NS5a\_1 \\
    & $15.20$	& PF03506 Flu\_C\_NS1      & $441.03$	& PF08301 HCV\_NS5a\_1b \\
    & $15.26$	& PF06593 RBDV\_coat       & $483.96$	& PF01506 HCV\_NS5a \\ \hline 
    
    & $0.01$	& PF00506 Flu\_NP    & $0.03$	& PF00096 zf-C2H2 \\
    & $0.01$	& PF00516 GP120      & $0.04$	& PF00078 RVT\_1 \\
    & $0.01$	& PF00540 Gag\_p17   & $0.17$	& PF00023 Ank \\
    & $0.01$	& PF00603 Flu\_PA    & $0.17$	& PF00589 Phage\_integrase \\
    & $0.01$	& PF00695 vMSA       & $0.19$	& PF00903 Glyoxalase \\
    $0.650 $ $\to$ $ 0.625$ & & & & \\
    & $12.57$	& PF06952 PsiA       & $202.08$	& PF01002 Flavi\_NS2B \\
    & $13.73$	& PF06788 UPF0257    & $221.93$	& PF01349 Flavi\_NS4B \\
    & $14.79$	& PF05788 Orbi\_VP1  & $222.59$	& PF01353 GFP \\
    & $15.42$	& PF00901 Orbi\_VP5  & $229.23$	& PF01350 Flavi\_NS4A \\
    & $16.02$	& PF03753 HHV6-IE    & $243.38$	& PF00948 Flavi\_NS1 \\ \hline
    
 \end{tabular}  
  \end{center} 
\end{table*}

\begin{table*}[!htb] 
  \begin{center}
    \begin{tabular}{|c|cc|cc|}  \hline 
   
    & $0.01$	& PF00124 Photo\_RC         & $0.09$  & PF00009 GTP\_EFTU \\
    & $0.01$	& PF00603 Flu\_PA           & $0.13$  & PF07974 EGF\_2 \\
    & $0.01$	& PF00695 vMSA              & $0.2$   & PF00096 zf-C2H2 \\
    & $0.01$	& PF01560 HCV\_NS1          & $0.22$  & PF00560 LRR\_1 \\
    & $0.02$	& PF00223 PsaA\_PsaB        & $0.23$  & PF01546 Peptidase\_M20 \\
    $0.625$ $\to$ $0.600$ & & & & \\
    & $11.95$	& PF06517 Orthopox\_A43R    & $376.41$ & PF01002 Flavi\_NS2B \\
    & $12.09$	& PF00843 Arena\_nucleocap  & $403.70$ & PF00948 Flavi\_NS1 \\
    & $13.08$	& PF06802 DUF1231           & $411.72$ & PF01349 Flavi\_NS4B \\
    & $14.72$	& PF05273 Pox\_RNA\_Pol\_22 & $425.27$ & PF01350 Flavi\_NS4A \\
    & $16.90$	& PF03021 CM2               & $538.21$ & PF05408 Peptidase\_C28 \\ \hline
    
    & $0.01$	& PF00517 GP41             & $0.06$	& PF00096 zf-C2H2 \\
    & $0.01$	& PF00559 Vif              & $0.06$	& PF00097 zf-C3HC4 \\ 
    & $0.01$	& PF00600 Flu\_NS1         & $0.09$	& PF00009 GTP\_EFTU \\
    & $0.01$	& PF00969 MHC\_II\_beta    & $0.09$	& PF01266 DAO \\
    & $0.01$	& PF06815 RVT\_connect     & $0.11$	& PF01926 MMR\_HSR1 \\
    $0.600$ $\to$ $0.575$ & & & & \\
    & $10.54$	& PF02477 Nairo\_nucleo    & $133.87$ & PF05790 C2-set \\
    & $11.95$	& PF07982 Herpes\_UL74     & $139.12$ & PF01353 GFP \\
    & $12.30$	& PF06871 TraH\_2          & $150.11$ & PF00518 E6 \\
    & $14.14$	& PF02509 Rota\_NS35       & $195.29$ & PF02929 Bgal\_small\_N \\
    & $16.04$	& PF06929 Rotavirus\_VP3   & $231.71$ & PF01382 Avidin \\ \hline
    
    & $0.01$	& PF00016 RuBisCO\_large   & $0.02$	& PF00115 COX1 \\
    & $0.01$	& PF00113 Enolase\_C       & $0.07$	& PF07690 MFS\_1 \\
    & $0.01$	& PF00123 Hormone\_2       & $0.08$	& PF07993 NAD\_binding\_4 \\
    & $0.01$	& PF00506 Flu\_NP          & $0.09$	& PF00517 GP41 \\
    & $0.01$	& PF01010 Oxidored\_q1\_C  & $0.10$	& PF00583 Acetyltransf\_1 \\
    $0.575 $ $\to$ $ 0.550$ & & & & \\ 
    & $10.60$	& PF06134 RhaA             & $161.43$ & PF01140 Gag\_MA \\
    & $10.95$	& PF07095 IgaA             & $168.19$ & PF04528 Adeno\_E4\_34 \\
    & $11.75$	& PF00897 Orbi\_VP7        & $173.44$ & PF08377 MAP2\_projctn \\
    & $12.13$	& PF03294 Pox\_Rap94       & $184.23$ & PF02093 Gag\_p30 \\
    & $13.75$	& PF01295 Adenylate\_cycl  & $311.32$ & PF01141 Gag\_p12 \\ \hline
    
    & $0.01$	& PF00016 RuBisCO\_large  & $0.06$	& PF00067 p450 \\
    & $0.01$	& PF00516 GP120           & $0.07$	& PF00023 Ank \\
    & $0.01$	& PF00522 VPR             & $0.08$	& PF00097 zf-C3HC4 \\
    & $0.01$	& PF00540 Gag\_p17        & $0.11$	& PF01381 HTH\_3 \\
    & $0.01$	& PF01539 HCV\_env        & $0.11$	& PF04851 ResIII \\
    $0.550$ $\to$ $0.525$             &       &                         &             &  \\
    & $11.29$	& PF05928 Zea\_mays\_MuDR & $101.41$	& PF01537 Herpes\_glycop\_D \\
    & $11.62$	& PF06829 DUF1238         & $121.18$	& PF02929 Bgal\_small\_N \\
    & $11.63$	& PF03277 Herpes\_UL4     & $123.25$	& PF01376 Enterotoxin\_b \\
    & $11.64$	& PF03395 Pox\_P4A        & $128.24$	& PF06466 PCAF\_N \\
    & $12.73$	& PF08405 Calici\_PP\_N   & $147.36$	& PF05806 Noggin \\ \hline

    & $0.01$	& PF00600 Flu\_NS1           & $0.02$	& PF00106 adh\_short \\
    & $0.01$	& PF00869 Flavi\_glycoprot   & $0.04$	& PF00270 DEAD \\
    & $0.01$	& PF01539 HCV\_env           & $0.05$	& PF00037 Fer4 \\
    & $0.01$	& PF02461 AMO                & $0.06$	& PF02518 HATPase\_c \\
    & $0.01$	& PF02788 RuBisCO\_large\_N  & $0.08$	& PF00249 Myb\_DNA-binding \\
    $0.525$ $\to$ $0.500$	&     &                           &        & \\
    & $11.36$	& PF07434 CblD              & $68.92$	& PF03939 Ribosomal\_L23eN \\
    & $11.80$	& PF04913 Baculo\_Y142      & $72.11$	& PF06267 DUF1028 \\
    & $11.98$	& PF05880 Fiji\_64\_capsid  & $96.66$	& PF02022 Integrase\_Zn \\
    & $13.48$	& PF06306 CgtA              & $120.34$	& PF00552 Integrase \\
    & $13.98$	& PF03317 ELF               & $129.98$	& PF02929 Bgal\_small\_N \\ \hline
  \end{tabular}   
\caption{\label{tab6} \small For proteins joining clusters outside the
  largest component or joining the giant component the five mostly
  underrepresented and five mostly overrepresented PFAM domains are
  giver per interval of weight w.}
 \end{center}
\end{table*}

\end{document}